\documentclass[twocolumn]{aastex63}

\usepackage{ulem}
\usepackage{hyperref}
\usepackage{amsmath}

\defcitealias{2008A&A...487..131C}{C08}
\defcitealias{2021AJ....161...79M}{M21}
\defcitealias{2015MNRAS.447.3973J}{J15}
\defcitealias{2011MNRAS.411..263J}{J11}
\defcitealias{2012ApJ...759..146M}{M12}

\shorttitle{Distances to M33}
\shortauthors{Lee et al.}
\graphicspath{{./}{figures/}}

\begin{document}

\title{The Astrophysical Distance Scale: V.\\ A 2\% Distance to the Local Group Spiral M33 via \\the JAGB Method, Tip of the Red Giant Branch, and Leavitt Law}

\author{Abigail~J.~Lee}\altaffiliation{LSSTC DSFP Fellow}\affiliation{Department of Astronomy \& Astrophysics, University of Chicago, 5640 South Ellis Avenue, Chicago, IL 60637}\affiliation{Kavli Institute for Cosmological Physics, University of Chicago,  5640 South Ellis Avenue, Chicago, IL 60637}

\author{Laurie~Rousseau-Nepton}\affil{Canada-France-Hawaii Telescope, Kamuela, HI, United States}

\author{Wendy~L.~Freedman}\affil{Department of Astronomy \& Astrophysics, University of Chicago, 5640 South Ellis Avenue, Chicago, IL 60637}\affiliation{Kavli Institute for Cosmological Physics, University of Chicago,  5640 South Ellis Avenue, Chicago, IL 60637}

\author{Barry~F.~Madore}\affil{Observatories of the Carnegie Institution for Science 813 Santa Barbara St., Pasadena, CA~91101}\affil{Department of Astronomy \& Astrophysics, University of Chicago, 5640 South Ellis Avenue, Chicago, IL 60637}

\author{Maria-Rosa~L.~Cioni}\affil{Leibniz-Institut für Astrophysik Potsdam (AIP), An der Sternwarte 16, D-14482 Potsdam, Germany}

\author{Taylor~J.~Hoyt}\affil{Department of Astronomy \& Astrophysics, University of Chicago, 5640 South Ellis Avenue, Chicago, IL 60637}\affiliation{Kavli Institute for Cosmological Physics, University of Chicago,  5640 South Ellis Avenue, Chicago, IL 60637}

\author{In Sung Jang}\affil{Department of Astronomy \& Astrophysics, University of Chicago, 5640 South Ellis Avenue, Chicago, IL 60637}\affiliation{Kavli Institute for Cosmological Physics, University of Chicago,  5640 South Ellis Avenue, Chicago, IL 60637}

\author{Atefeh~Javadi}\affil{School of Astronomy, Institute for Research in Fundamental Sciences (IPM), P.O. Box 1956836613, Tehran, Iran}

\author{Kayla~A.~Owens}\affil{Department of Astronomy \& Astrophysics, University of Chicago, 5640 South Ellis Avenue, Chicago, IL 60637}\affiliation{Kavli Institute for Cosmological Physics, University of Chicago,  5640 South Ellis Avenue, Chicago, IL 60637}

\correspondingauthor{Abigail J. Lee}\email{abbyl@uchicago.edu}

\begin{abstract}
The J-region asymptotic giant branch (JAGB) method is a new standard candle that is based on the stable intrinsic $J$-band magnitude of color-selected carbon stars,  and has a precision comparable to other primary distance indicators such as Cepheids and the TRGB. We further test the accuracy of the JAGB method in the Local Group Galaxy M33. M33's moderate inclination, low metallicity, and nearby proximity make it an ideal laboratory for tests of systematics in local distance indicators. Using high-precision optical $BVI$ and near-infrared $JHK$ photometry, we explore the application of three independent distance indicators: the JAGB method, the Cepheid Leavitt Law, and the TRGB. We find: 

$$\mu_0 (TRGB_I) = 24.72 \pm 0.02 \textrm{(stat)} \pm0.07 \textrm{(sys)~mag},$$
$$\mu_0 (TRGB_{NIR}) = 24.72 \pm 0.04 \textrm{(stat)} \pm0.10 \textrm{(sys)~mag},$$
$$\mu_0 (JAGB) = 24.67 \pm 0.03 \textrm{(stat)} \pm0.04 \textrm{(sys)~mag},$$ 
$$\mu_0 (Cepheid) = 24.71 \pm 0.04 \textrm{(stat)} \pm0.01 \textrm{(sys)~mag}.$$

For the first time, we also directly compare a JAGB distance using ground-based and space-based photometry. We measure:
$$\mu_0 (JAGB_{F110W}) = 24.71 \pm 0.06 \textrm{(stat)} \pm0.05 \textrm{(sys)~mag}$$
using the (F814W$-$F110W) color combination to effectively isolate the JAGB stars. 

In this paper, we measure a distance to M33 accurate to 2\% and provide further evidence that the JAGB method is a powerful extragalactic distance indicator that can effectively probe a local measurement of the Hubble constant using spaced-based observations. We expect to measure the Hubble constant via the JAGB method in the near future, using observations from JWST.
\end{abstract}

\keywords{Cepheid distance (217), Hubble constant (758), Observational cosmology (1146), Distance indicators (394), Population II stars (1284), Asymptotic Giant Branch stars (2100), Carbon stars (199), Cepheid variable stars (218), Red giant tip (1371), Galaxy distances (590), Triangulum Galaxy (1712)}

\section{Introduction}

Within the past decade, a distinct 10\% difference has arisen between the increasingly precise measurements of the Hubble constant inferred from fits of the standard $\Lambda$ cold dark matter ($\Lambda$CDM) model to the measured power spectrum of cosmic microwave background (CMB) anisotropies (e.g. \citealt{2020A&A...641A...6P, 2020JCAP...12..047A}) and local type Ia supernovae (SNe Ia) distance ladders calibrated by Cepheids (e.g. Hubble Space Telescope (HST) Key Project: \citealt{2001ApJ...553...47F, 2012ApJ...758...24F}; Supernovae and $H_0$ for the Equation of State of dark energy (SHoES) group: \citealt{2021ApJ...908L...6R}).
While this  agreement is actually quite extraordinary, given that the two independent measurements span from the early universe to billions of years later today, the value of the Hubble constant is so intrinsically covariant with other cosmological parameters that it remains critical that we understand the divergence.

The 5-$\sigma$ tension points to either unknown systematics affecting any or all of the probes, \citep{2021arXiv210301183D, 2021Ap&SS.366...99S} or a necessary extension to the $\Lambda$CDM model.
However,  using the Tip of the Red Giant Branch (TRGB) as the SN Ia calibrator, \cite{2021arXiv210615656F} recently found no statistically significant difference between the values of $H_0$ measured locally and inferred from observations of the CMB, and only a 2-$\sigma$ tension with the SHoES group \citep{2016ApJ...826...56R, 2019ApJ...876...85R, 2021ApJ...908L...6R}. 
The J-region Asymptotic Giant Branch (JAGB) method, a standard candle in the near-infrared, can provide not only a third independent calibration of SNe Ia for a measurement of $H_0$, but can also serve as a cross-check for Cepheid and TRGB-based distances. These types of independent evaluations are a pivotal step toward understanding and then resolving the Hubble tension.

In earlier papers of this series, we introduced the JAGB method and determined a 
provisional zero point (Paper I: \citealt{2020arXiv200510792M}); we then applied
the method to 14 galaxies whose distances range from 50 kpc to 4 Mpc  (Paper II: \citealt{2020ApJ...891...57F}). Despite the fact that all of the data for the JAGB stars in those galaxies were obtained using a variety of telescopes and for different scientific objectives, a comparison of the JAGB distances 
with the TRGB distances to those same galaxies found agreement in the mean to within $\pm0.025$~mag (using $M_J = -6.20$~mag for the JAGB, \citealt{2020arXiv200510793F}; and $M_I =$~$-4.05$~mag for the TRGB, \citealt{2020ApJ...891...57F}). Furthermore, the inter-sample scatter found from the JAGB$-$TRGB comparison also indicated 
that the precision of either method must be better than  $\pm$0.05~mag.
The two methods have little in common with respect to known 
systematics, given that they are drawn from two entirely different stellar 
populations. The TRGB stars are the oldest, lowest-mass, low-metallicity 
Population~II stars, and are deliberately sampled only in the outer gas- 
and dust-free halos of their parent galaxies where crowding is minimized by experimental design. In contrast, the JAGB stars are intermediate-age,
intermediate-mass stars and are found in the thick/extended disks of galaxies with 
recent (but not necessarily ongoing) star formation. 
JAGB stars are also prolific 
enough that large samples can be obtained in the outermost regions of the 
extended disks, minimizing both reddening and crowding. 
Furthermore, the small observed inter-sample scatter in the distance moduli calculated with the JAGB method with respect to the TRGB method must include differences resulting from host galaxy type, metallicities, and star formation histories, hinting that none of these effects produce significant systematics hindering the JAGB method. In Paper III \citep{2021ApJ...907..112L}, we further confirmed the accuracy and precision of the JAGB method. In that case we measured and compared the distances to the Local Group galaxy WLM using the Cepheid Leavitt Law, near-infrared and optical TRGB, and JAGB method, finding better than 3\% agreement. In Paper IV \citep{jagbpaper4}, we provided a provisional JAGB zeropoint in the Hubble Space Telescope WFC3/IR F110W filter. 

With this paper we continue the process of uniformly observing all highly-resolvable galaxies within reach of
ground-based telescopes in the near-infrared, obtaining photometry of the JAGB stars in the least-crowded regions in each of these galaxies. We also test the HST zeropoint determined in Paper IV by comparing the JAGB distance modulus obtained from ground-based imaging using the conventional $J$ band with ($J-K$) color cuts, to that obtained from space-based imaging using the F110W filter with (F814W$-$F110W) color cuts. To date, a direct comparison of the two has been unexplored, and it will be a valuable test for the accuracy of future space-based JAGB distance determinations.

We start here with M33 (NGC 598), a nearby spiral galaxy of type Sc II-III and the third brightest member of the Local Group. M33's low metallicity, minimal reddening, and moderate inclination of $56^{\circ}$ \citep{1989AJ.....97...97Z} make it an ideal laboratory for testing distance indicators. 
The distance to M33 was first measured by
\cite{1926ApJ....63..236H}, and was one of the first galaxies he measured using Cepheids as distance indicators.
The first effort at using TRGB stars to measure distances in M33 came more than half a century later from \cite{1986ApJ...305..591M}, who utilized the observed brightness of the tip of the giant branch as a standard candle.
And finally, about 15 years ago \cite{2005AJ....129..729R} found the carbon star luminosity function (LF) in M33 to be similar to that of M31 and the SMC, ``suggesting that C stars should be useful distance indicators.'' These three distinct types of stars (Cepheids, TRGB stars, and carbon stars) and their refined
methods for determining absolute stellar magnitudes each have different strengths 
and largely independent systematics. Together they are three equally solid, interlocking
pillars, establishing the basis of an {\it Astrophysical Distance Scale.}

The outline of this paper is as follows. In Section \ref{sec:phot} we describe the photometry utilized in this study. 
In Section \ref{sec:jagb}, we present our JAGB measurements using both ground-based and space-based HST data.
In Section \ref{sec:trgb}, we give an overview of the multi-wavelength TRGB method and apply it to M33. 
In Section \ref{sec:ceph}, we identify Cepheids previously published in the literature and measure a Leavitt law distance to M33.
In Section \ref{sec:compre}, we compare our measured distances to M33 with distances in the literature. And finally in Section \ref{sec:sum}, we give a summary of this work and look to the future.

\section{Multi-wavelength Observations of M33}\label{sec:phot}

\subsection{Optical BVI Photometry}\label{subsec:optphot}

Our optical $BVI$ photometry was obtained from and described in detail in \cite{2009MNRAS.396.1287S}, used for the purposes in this study for measuring the I-band TRGB and a multi-wavelength Leavitt law distance.  Imaging was obtained between 1998 and 2001 on the 3.5-m WIYN telescope using the S2KB imager and Mini-Mosaic camera, which have resolutions of 0.195 and 0.141 arsec/pixel, respectively. 
This imaging study targeted Cepheids in the inner and outer regions of the galaxy, with the goal of quantifying metallicity effects on the Cepheid period-luminosity relation.

The TRGB is best measured in the outer, less crowded regions of galaxies where effects of population blending, reddening, and metallicity are decreased. Furthermore, several studies have found significant differences between distance moduli measured from Cepheids in the outer and inner regions of M33. \cite{2009MNRAS.396.1287S} found differences on the order of 0.17~mag between reddening-free distance moduli measured by Cepheids in the inner and outer regions of M33, respectively, but whether this difference is due to blending/crowding or metallicity is still debated \citep{2012AJ....144..113C, 2013ApJ...773...69G}.
Thus, we utilized the \citealt{2009MNRAS.396.1287S} ``outer field'' data set in the southern part of the galaxy, which targeted the sparser metal-poor populations.

\subsection{Near-Infared JHK Photometry}

\begin{figure}
\centering
\includegraphics[width=\columnwidth]{"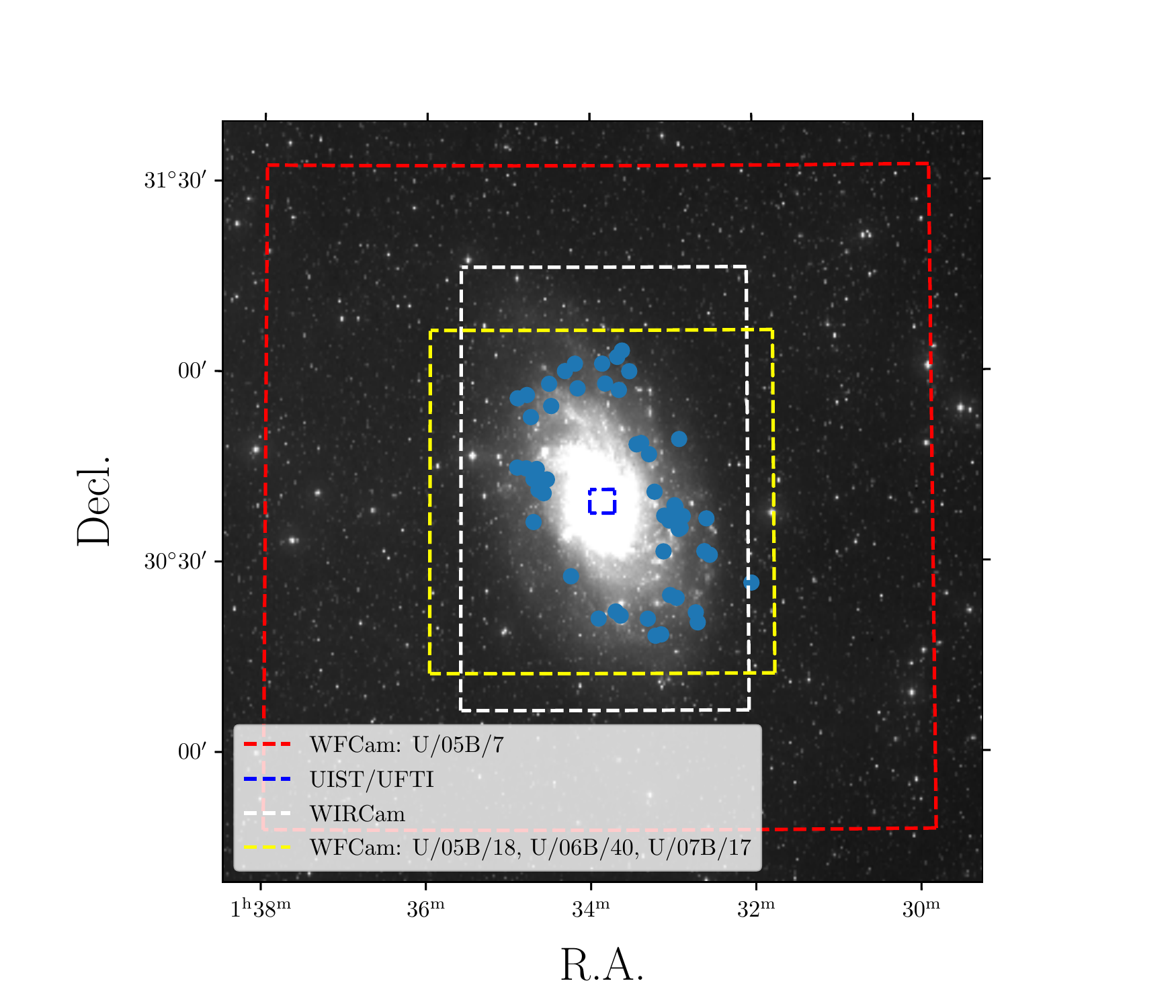"}
\caption{Digitized Sky Survey image of the Local Group galaxy M33. The outlines of the regions covered in the four near-infrared catalogs used to measure the JAGB are marked. The locations of the 60 Cepheids used in this study are shown as blue dots, described in Section \ref{sec:ceph}. }
\label{fig:f1}
\end{figure}

While JAGB stars are indeed variable with an intrinsic dispersion in their population mean of $\pm0.2$~mag \citep{2001ApJ...548..712W}, averaging over multiple epochs of data decreases the intrinsic scatter of their observed $J$-band luminosity function as $\pm 1/\sqrt{N_{obs}}$ mag, where $N_{obs}$ is the number of observations.  
Therefore, to reduce the observed scatter, we utilized four near-infrared JHK imaging data sets in this study, from: (1) the Wide-Field CAMera (WFCam) on the 3.8 m UK InfraRed Telescope  (UKIRT; previously published by \citealt{2008A&A...487..131C} and further post-processed by \citealt{2021AJ....161...79M}), (2)  UKIRT/WFCam (previously published by \citealt{2015MNRAS.447.3973J}), (3) the UIST/UFTI imagers on UKIRT (previously published by \citealt{2011MNRAS.411..263J}), and (4) the Wide-Field InfraRed CAMera (WIRCam) on the 3.6 m Canada-France-Hawaii Telescope (CFHT; published here for the first time).
A visual representation of the areas targeted in each of the four catalogs is shown in Figure \ref{fig:f1}. We discuss each of these data sets in turn below.

\subsubsection{WFCam: U/05B/7}\label{subsubsec:wfcam}
The first dataset, described in \citealt{2008A&A...487..131C} (hereafter \citetalias{2008A&A...487..131C}), used the WFCam on the UKIRT in 2005 under program U/05B/7 to survey the luminous red stellar populations of M33. 
These observations were part of a program to probe the near-infrared properties of Local Group galaxies \citep{2013ASSP...37..229I}.

The WFCam makes use of four 2048$\times$2048 pixel detectors arranged in a 2$\times$2 pattern. Each detector is spaced apart by approximately 90\% of a detector width in both axes. By combining four stepped exposures into one mosaic, an area (tile) of approximately 0.75 deg$^2$ is covered. The pixels are 0.4 arcsec in size. 
This dataset consists of four tiles covering the central 3 deg$^2$ of M33. Each tile was observed in total for 150s in J and 270s in H and K, including multiple slightly shifted (jitter and/or microstepping) exposures. 
The average seeing during the observations was 1.07$\pm$0.06 arcsec. The data were reduced using the WFCam pipeline, which provides PSF photometry, and then calibrated with respect to 2MASS. 
The JK photometry was then further cleaned and combined in \citealt{2021AJ....161...79M} (hereafter \citetalias{2021AJ....161...79M}) for a study on the red supergiants in M33, and kindly provided to us by Philip Massey. The final photometry list included 121,328 stellar objects.

\subsubsection{WFCam: U/05B/18, U/06B/40, U/07B/17}

The next dataset, described in more detail in \citealt{2015MNRAS.447.3973J} (hereafter \citetalias{2015MNRAS.447.3973J}), also utilized WFCam at UKIRT under observing programs U/05B/18, U/06B/40, and U/07B/17. Observations of M33 were obtained from September 2005 to October 2007 to study variable red giant stars in the disk of M33. The data were reduced and calibrated to the 2MASS photometric using the WFCam pipeline. Their catalog originally contained 403,733 stars. We cleaned the catalog using the photometric $J$-band uncertainty, using the following constant+exponential function for the photometric uncertainty $\sigma_J$ as a function of the $J$-band magnitude, consistent with CCHP procedure (e.g., see the appendix of \citealt{2019ApJ...885..141B}):

\begin{equation}
    \sigma_J < 0.035 + 0.003 \times e^{m_J-16.7}.
\end{equation}

The final cleaned catalog contained 129,028 stars. 

\subsubsection{UIST/UFTI}
The third dataset we utilized is described in \citealt{2011MNRAS.411..263J}  (hereafter \citetalias{2011MNRAS.411..263J}), and was obtained to also study the variable red giants in M33. Observations were taken from October 2003 to July 2007, using UKIRT's UIST and UFTI imagers. Photometry was obtained using DAOPHOT \citep{1987PASP...99..191S}. Non-stellar objects were removed based on their $\chi$ value in \citetalias{2011MNRAS.411..263J}, with the final catalog having 18,398 objects. The photometry had also been previously transformed to the 2MASS photometric system using transformations from \cite{2001AJ....121.2851C}.

\subsubsection{WIRCam}

The fourth catalog was extracted from the observing program 18BD94 (PI: Rousseau-Nepton), that used the WIRCam on the CFHT for infrared follow up of the SIGNALS survey of nearby galaxies \citep{ 2019MNRAS.489.5530R}. This dataset consisted of a mosaic of 6 fields covering the disk of M33 using the dithering pattern \textit{DP20} with 20 positions and a total exposure time averaging 23m20s, 33m20s and 35m00s for J, H and K band respectively. Note that overlaps due to the mosaic nature of the observations increased the signals in some of the area covered. 
Data were initially calibrated with standard stars' derived zero-points, and the image quality averaged between 0.5 and 0.9 arcsec. 
Astrometry, photometric variability corrections, and stacking were performed using the \textsc{AstrOmatic.net} suite of packages (\textsc{SExtractor}, \textsc{SCAMP}, \textsc{SWarp}: \citealt{1996A&AS..117..393B, 2006ASPC..351..112B, 2002ASPC..281..228B}). A catalog of all the sources was then extracted over the final stacked mosaic for each band using \textsc{SExtractor} again. For this extraction, we opted for an aperture of 12 WIRCam pixels (corresponding to 3.7 arcsec) and then used weight maps to recover properly the statistical significance of all sources over a detection threshold of 3-$\sigma$. The catalog contained initially 193,647 sources from which the following selection criteria were applied to keep only the stellar sources, after which 72,375 sources remained:

\begin{equation*}
J_{ELONG}, H_{ELONG}, K_{ELONG}<1.4
\end{equation*}
\begin{equation*}
J_{RADIUS}, H_{RADIUS}, K_{RADIUS}<2
\end{equation*}
\begin{equation*}
J_{FLAGS}, H_{FLAGS}, K_{FLAGS}=0
\end{equation*}

Then, we transformed our photometry to the 2MASS photometric system using equations provided by the WIRwolf image stacking pipeline.\footnote{\url{http://www.cadc-ccda.hia-iha.nrc-cnrc.gc.ca/en/wirwolf/docs/filt.html} }

\begin{equation*}
J_{2MASS} = J_{WIRCam} + 0.071 \times  (J_{WIRCam}-H_{WIRCam})
\end{equation*}
\begin{equation*}
H_{2MASS} = H_{WIRCam}  - 0.034 \times  (J_{WIRCam}-H_{WIRCam})
\end{equation*}
\begin{equation*}
    \begin{aligned}
K_{2MASS} = K_{WIRCam} - 0.062 \times   (H_{WIRCam}-K_{WIRCam}) + \\
0.002 \times (J_{WIRCam} - H_{WIRCam})
    \end{aligned}
\end{equation*}

\subsubsection{Merging the Catalogs}\label{subsubsec:merge}
\begin{figure*}
\centering
\includegraphics[width=\textwidth]{"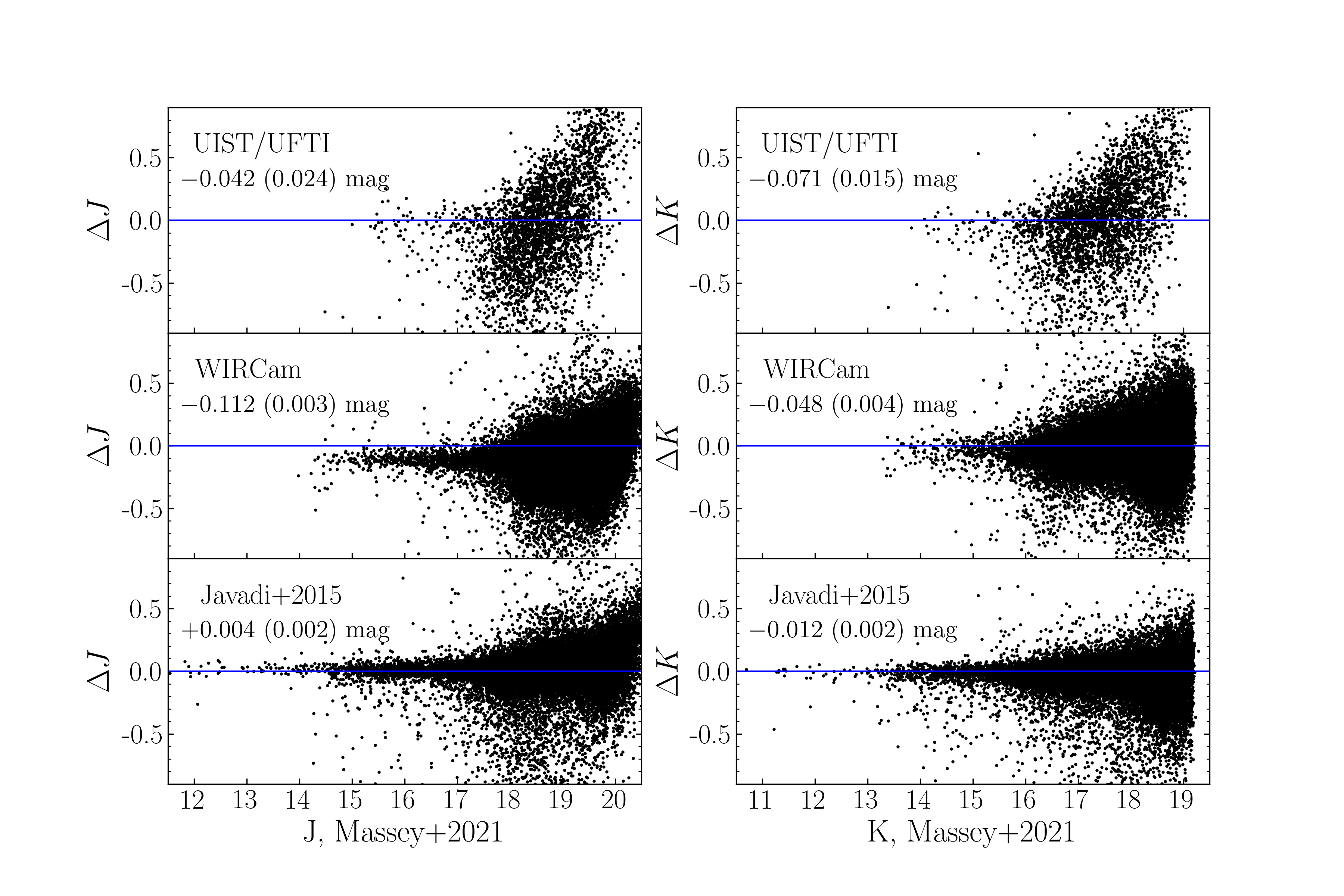"}
\caption{Comparison of photometry in the $J$ and $K$ bands between the  \citetalias{2008A&A...487..131C}/\citetalias{2021AJ....161...79M} WFCam catalog and: (top) \citetalias{2011MNRAS.411..263J} UIST/UFTI catalog, (middle) WIRCam catalog, and (bottom) WFCam catalog from \citetalias{2015MNRAS.447.3973J}. The median magnitude offsets and errors on the mean are labeled in each panel. $\Delta J$ and $\Delta K$ are defined as ``\citetalias{2021AJ....161...79M} $-$ catalog.''}
\label{fig:f16}
\end{figure*}

To merge the four catalogs, we first measured the $JK$ photometric offsets with respect to the \citetalias{2021AJ....161...79M} catalog, as it had been the most extensively calibrated to 2MASS. To derive the offsets, we first matched the four catalogs using a matching radius of $1\arcsec$ in \textsc{Topcat}. 
We then compared the photometry of the bright sources in the magnitude range $15<J<16.5$~mag.\footnote{In Figure \ref{fig:f16}, it appears as if there are magnitude dependent offsets for the fainter stars. This results from the well-documented fact that sources near to the detection threshold of a photometric survey are less likely to be found in a comparison between two different photometric systems, especially if one survey has a brighter detection threshold  \citep{1940MNRAS.100..354E}. Therefore, we chose to use the bright stars to calibrate our photometry, rather than the stars in the TRGB or JAGB star magnitude ranges, to avoid any biases incurred by incompleteness at faint magnitudes. The bright stars showed no evidence of any magnitude dependent offsets.}
We calculated the median photometric offsets with respect to  \citetalias{2021AJ....161...79M}, rejecting stars that deviated more than 2-$\sigma$ from the median.
As expected, the  \citetalias{2015MNRAS.447.3973J} and \citetalias{2011MNRAS.411..263J}  catalogs showed only small offsets as they had previously been nominally calibrated to 2MASS, whereas the WIRCam catalog needed slightly larger corrections. We then applied the determined photometric corrections to the three respective catalogs. All of the derived offsets and photometry comparisons are shown in Figure \ref{fig:f16}.
Finally, we merged the four catalogs; if a star was found in two or more catalogs, we adopted the weighted mean magnitude. If it was only found in one, we kept that single magnitude point. Photometric uncertainties were also updated for the mean magnitudes (i.e. $\sigma = \sqrt{\sum_{i=1}^{N} \sigma_i^2} /N$).

\subsubsection{H-band Photometry}
The primary purpose of this paper was to measure a JAGB method distance modulus to M33, for which we only need $JK$-band photometry. However, we also measured TRGB and Leavitt law distances as a cross-check. To measure $JHK$ TRGB and multi-wavelength Leavitt law distances, we also needed to create a merged H-band catalog. In section \ref{subsubsec:merge}, we treated the \citetalias{2021AJ....161...79M} as fiducial, as it had been subject to the most cleaning and post-processing. However, the \citetalias{2021AJ....161...79M} catalog did not  post-process the H-band photometry, so we instead treated the original \citetalias{2008A&A...487..131C} H-band photometry as fiducial (which was obtained from the same pipeline and observations as the \citetalias{2021AJ....161...79M} data).

\begin{figure}
\centering
\includegraphics[width=\columnwidth]{"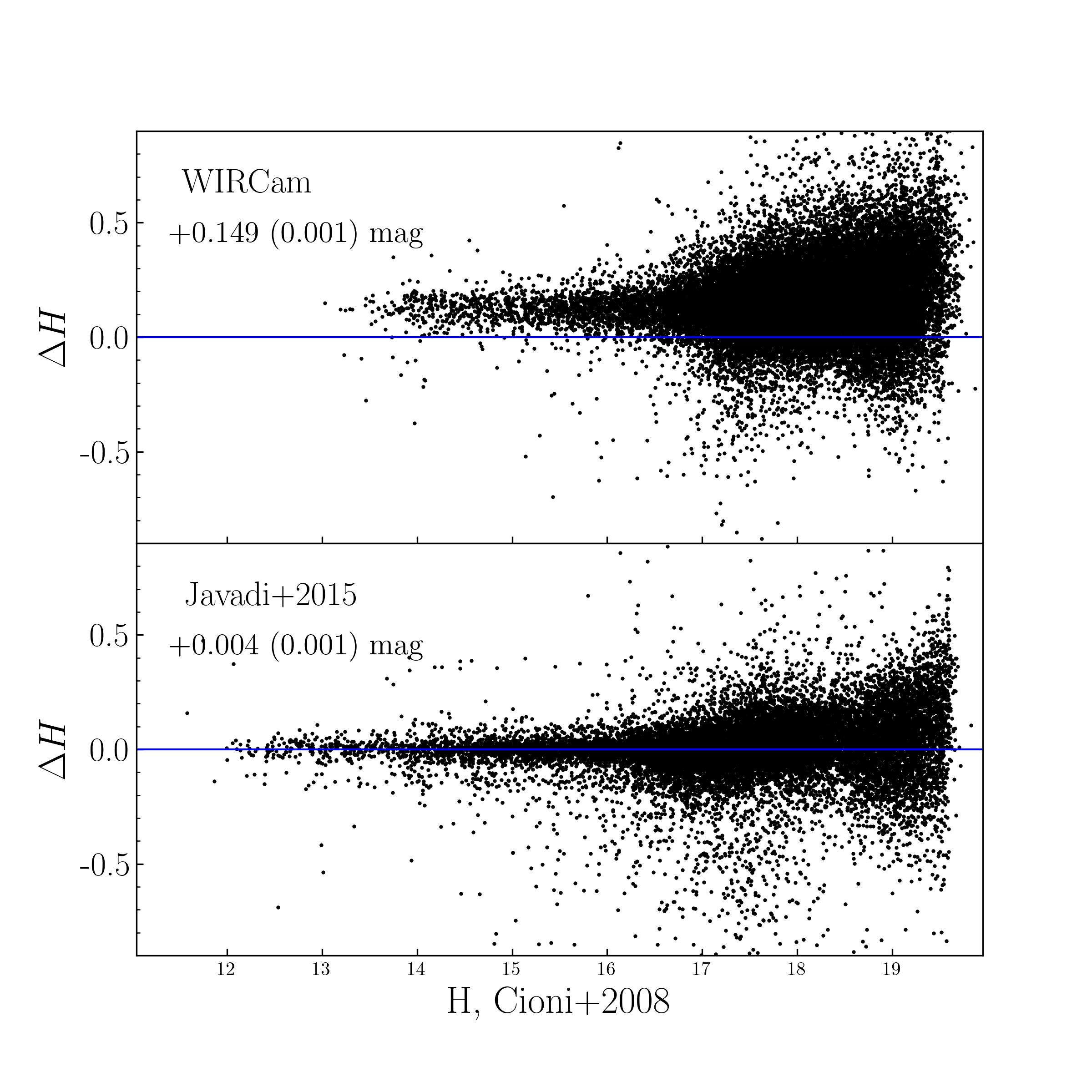"}
\caption{Comparison of photometry in the H band between the \citetalias{2008A&A...487..131C} catalog and \citetalias{2015MNRAS.447.3973J} and WIRCam catalogs, used for measuring the H-band TRGB. The median offsets and errors on the mean are labeled in each panel. $\Delta H$ is defined as ``\citetalias{2008A&A...487..131C} $-$ catalog.''  }
\label{fig:f5}
\end{figure}

We merged the \citetalias{2008A&A...487..131C},
\citetalias{2015MNRAS.447.3973J}, and WIRCam catalogs (the \citetalias{2011MNRAS.411..263J} was located in the crowded inner disk and thus unsuitable for our TRGB and Leavitt law measurements). 
Using the same cleaning/calibration procedure from \citetalias{2021AJ....161...79M}, we only selected objects from the \citetalias{2008A&A...487..131C} catalog classified as ``stellar'' or ``marginally stellar'' in the H band by the standard Cambridge Astronomy
Survey Unit (CASU) pipeline \citep{2004SPIE.5493..411I}, which uses the star's curve of growth to calculate a stellarness-of-profile statistic. Next, a comparison of the H-band source list with that of 2MASS showed an additional correction of +0.027~mag was needed. 
We then derived the additional H-band offsets between the \citetalias{2008A&A...487..131C} catalog and the \citetalias{2015MNRAS.447.3973J}  WFCam and WIRCam catalogs as in Section \ref{subsubsec:merge}. The photometry comparisons and derived offsets are shown in Figure \ref{fig:f5}. Finally, we applied the determined corrections to the two catalogs, and merged the three catalogs.

\subsection{HST Photometry}
To measure the JAGB magnitude in M33 using space-based photometry, we used  \textit{HST} photometry of M33 from the PHATTER survey \citep{2021ApJS..253...53W}, whose aim was to provide resolved stellar photometry for 22 million stars in the entire extended disk of M33. We utilized their WFC3/IR F110W (wide J-band equivalent) and ACS/WFC F814W (I-band equivalent) filters in order to identify the JAGB stars. In addition, we applied their ``good star" quality requirements in the F110W and F814W filters. To avoid the inner disk regions, we limited our photometry to three of their outer disk fields\footnote{See this link for a full visual of the bricks layout: \url{https://archive.stsci.edu/hlsp/phatter}} : the SE section of brick 3, and the NW and NN sections of brick 1.

\section{The JAGB Method}\label{sec:jagb}

JAGB stars are thermally-pulsating, intermediate-age AGB stars that have accumulated enough carbon in their atmospheres that they are red enough to be photometrically distinct in color-magnitude space. JAGB stars are double-shell burning, with helium and hydrogen shells that surround a degenerate carbon core. Their He-shell  produces carbon via triple-$\alpha$ reactions, and is subject to thermonuclear pulses \citep{1965ApJ...142..855S}. At the beginning of each pulse, the He-shell begins to burn strongly with luminosities up to $10^8 L_{\odot}$ \citep{2003agbs.conf.....H}. The resulting energy, which cannot be transported by radiation alone due to the opaqueness of the He-rich layers, creates a convective shell above the He-shell which lasts for a few hundred years. As the He-shell powers down, the convective shell shrinks and the outer convective envelope then penetrates deeper layers, bringing material up with it, 
With each thermal pulse, which happen every $10^3-10^5$~years, the H-rich convective envelope penetrates deeper into the star, until it reaches the C-rich layer after about $\sim 10-20$~thermal pulses (depending on the initial mass and molecular opacity of the star), in what is known as `the third dredge-up.' Carbon is brought up into the surface, thus giving rise to carbon stars \citep{2003agbs.conf.....H}

Twenty years ago \cite{2001ApJ...548..712W} set the wheels in motion for JAGB stars to be identified as powerful and precise distance indicators in the near infrared,  convincingly demonstrated by their 
ability to differentially map out the detailed 3-dimensional (back-to-front) geometry of the LMC by employing thousands of these stars distributed across the entire face of that galaxy. 
In the J band, the absolute peak luminosity of the JAGB luminosity function is remarkably stable throughout a variety of galaxy morphologies and inclinations, and empirically has been shown to be an effective standard candle \citep{2020arXiv200510793F}. In this section, we employ the JAGB method as a distance indicator in M33.

\subsection{Ground}\label{subsec:ground}

\begin{figure}
\centering
\includegraphics[width=\columnwidth]{"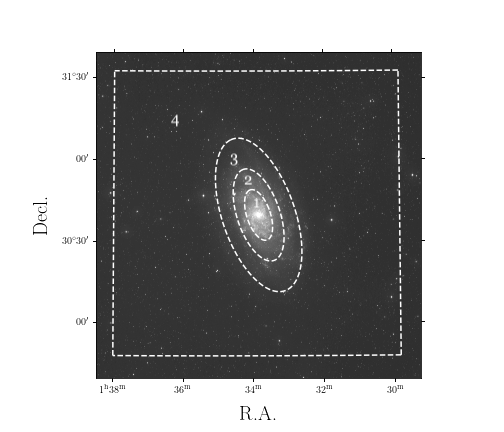"}
\caption{The regions delineated by concentric elliptical rings used in this study. Each region contains an approximately equal number of JAGB stars.}
\label{fig:fjagbspace}
\end{figure}

To assess potential reddening, population blending, and crowding effects, we measured the JAGB apparent magnitude in four regions delineated by concentric elliptical rings, each with approximately the same number of JAGB stars. The spatial selection is shown in Figure \ref{fig:fjagbspace}, where the ellipses have been centered on $(\alpha, \delta) = (23.46^{\circ}, 30.66^{\circ})$ with a position angle of the major axis of each ellipse of $23^{\circ}$ \citep{2000glg..book.....V}. Each region contained $\sim 6,860$ JAGB stars, a more than sufficient statistical sample. Region 1, the innermost region, is comprised of the inner disk of M33. Region 2, the next most innermost region, encompasses the rest of M33's inner disk. Region 3 encloses M33's extended outer disk and region 4 covers its halo\footnote{We nominally call the outer region "the halo," although the existence of the stellar halo around M33 is controversial \citep{2016MNRAS.461.4374M}.}.

\begin{figure*}
\centering
\includegraphics[width=\textwidth]{"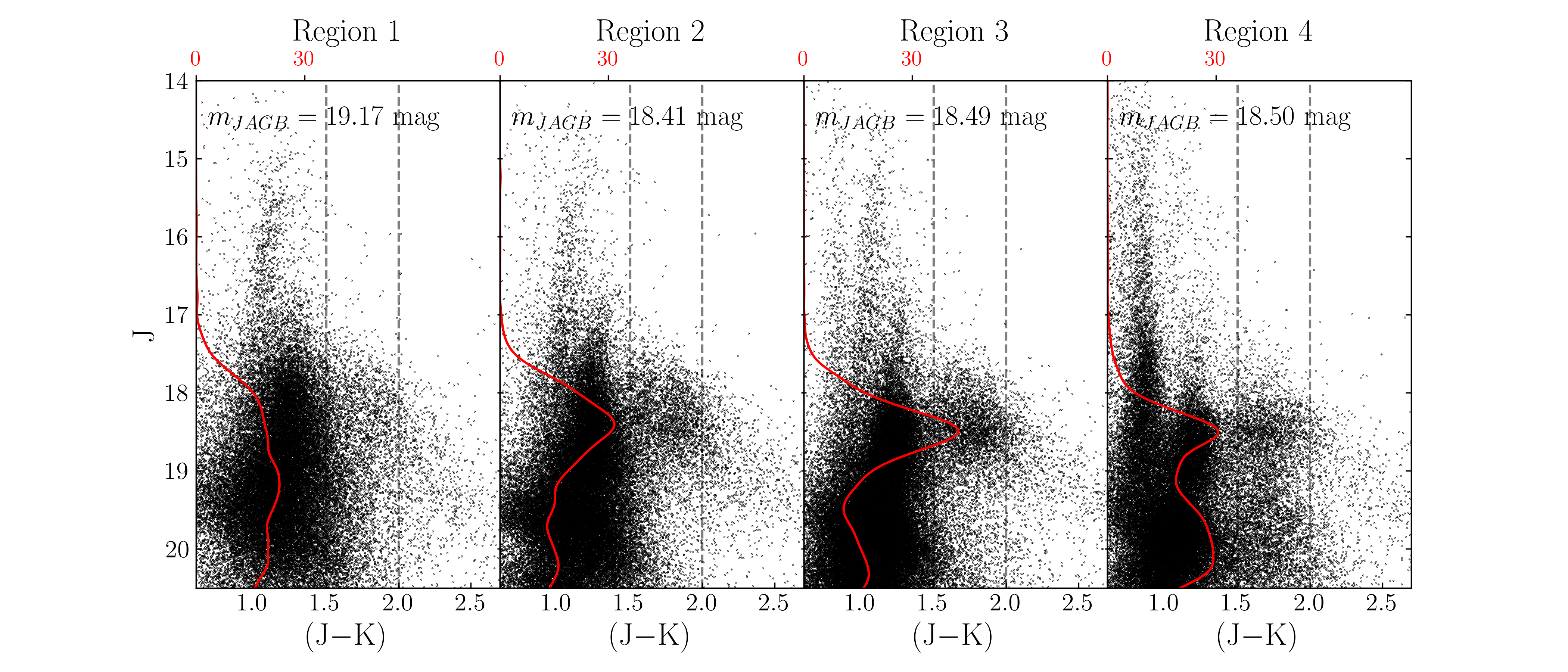"}
\caption{Color magnitude diagrams for the four regions. The JAGB stars are located in the color range $1.5<(J-K)<2.0$~mag. The smoothed luminosity functions for the JAGB stars are shown in red. The values for the peak of the JAGB luminosity function for each region is printed at the top of each panel.  The smoothing parameter $\sigma_s$ was chosen to be 0.22~mag for visualization purposes. 
\label{fig:f4}}
\end{figure*}

The color magnitude diagrams and luminosity functions for each region are shown in Figure \ref{fig:f4}, created from the merged $JK$ catalog.
The luminosity function was generated by first finely binning the J-band magnitudes over a color range of $1.5<(J-K)<2.0$~mag  with bins of 0.01~mag.\footnote{Traditionally, we have used a blue color cut of $(J-K)>1.4$~mag, but we found the M33 O-rich AGB population extends farther into the blue, so we instead adapted the color cut to 1.5. \cite{2021arXiv210502120Z} found the same in their study of JAGB stars in M33, choosing a color cut of $(J-K)>1.45$~mag. \citetalias{2008A&A...487..131C} also used $(J-K)>1.5$~mag as their blue color cut to delineate C and M AGB stars in M33. We plan to thoroughly investigate how the color selection of the JAGB stars changes in different environments in an exploration of the JAGB method in 13 nearby galaxies using observations obtained at the Magellan Telescope (Lee et al. in prep).} 
To control for Poisson noise in the  binned LF, we implemented the GLOESS (Gaussian-windowed, locally weighted scatterplot smoothing) algorithm.
GLOESS smoothing is a data-smoothing interpolating technique that is effective at suppressing false (noise-induced) edges and peaks in luminosity functions, especially in the case of low or empty bins. 
The GLOESS algorithm has been previously used in an astrophysical context for fitting variable-star light curves \citep{2004AJ....128.2239P, 2017AJ....153...96M} and for determining the position of the TRGB (e.g. \citealt{2017ApJ...845..146H}).

We note that other groups have used a slightly different procedure for measuring the apparent JAGB magnitude than that described above.
\cite{2020MNRAS.495.2858R} and \cite{2021MNRAS.501..933P} fit a modified Lorentzian distribution to the observed JAGB LF using a maximum likelihood estimator to measure the median magnitude. \cite{2021arXiv210502120Z} also fit a model to the observed JAGB LF, although using instead a Gaussian quadratic function. 
We chose rather to measure the apparent JAGB magnitude as the peak of the GLOESS-smoothed luminosity function, with the only user input being the smoothing parameter $\sigma_s$, which we discuss later in this section. We plan to further explore the accuracy of this method in a program to obtain high-precision JHK data for nearby galaxies, all obtained with the same telescope/instrument combination (Lee et al. in prep).

The smoothed luminosity function of the JAGB stars in each of the four regions provides some insight into effects due to  blending of other populations, crowding, and reddening. 
First, the peak of the JAGB luminosity function is the sharpest and clearest in the outer disk in region 3. Region 4 encompasses the halo of M33 and contains a high number of objects fainter than $J=18.5$~mag in the same color region of the JAGB stars (or $K\sim17$~mag), which \citetalias{2008A&A...487..131C} and  \citetalias{2021AJ....161...79M} both corroborate in their Figures 6 and 7 respectively as non-stellar sources (mostly blends but also some HII regions, stellar associations, and background galaxies). These extended sources were most prevalent in region 4 likely because it covered the most area, and there were fewer stars to mask the extended sources. 
Because we utilized archival photometry from other studies, most of the catalogs lacked the ``sharp'' or ``roundness'' parameters that are typically used to easily cull extended sources. For future studies using space-based data, we plan to perform our own PSF photometry, where the ``sharp'' and ``roundness'' parameters will be readily available from our own photometry pipelines. For a detailed discussion on the potential contamination from background galaxies on the JAGB magnitude, we refer the reader to an earlier companion paper, \cite{jagbpaper4}.
\cite{2005AJ....129..729R} also speculated that as the ratio of C to M stars begins to drop off outside of the outer disk, these objects may also be reddened M-type AGB stars.
Second, the JAGB star luminosity function in region 2 appears to be smeared out with less of a clearly defined peak, with region 1's LF even more so, likely from a confluence of higher crowding, blending, and reddening effects. The peak value of region 2's luminosity function is also the brightest of the four luminosity functions, potentially indicating issues with crowding, which can bias individual star magnitudes brighter. This is clear evidence that JAGB method measurements \textit{should avoid being made in the crowded inner disks} of galaxies. Even with high-precision multiple-epoch catalogs, the peak of the JAGB luminosity function is still not clearly defined in the crowded inner disk. 

Thus, to measure the JAGB apparent magnitude in M33, we used regions 3 and 4, which had a combined total of 13,701 JAGB stars. 
The peak magnitude of the smoothed luminosity function was measured to be $m_J=18.50\pm 0.01$~mag (error on the mean).
Changing the smoothing parameter from a range of $[0.10, 0.20]$ varied the JAGB apparent magnitude by at most 0.03~mag, which we thus adopted as a statistical uncertainty. 
To measure the true distance modulus, we used the absolute calibration from \cite{2020arXiv200510792M}: $-6.20\pm0.01$ (stat) $\pm0.04$ (sys)~mag.\footnote{The \cite{2020arXiv200510792M} calibration is set by the LMC and SMC geometric DEB distances. We chose not to utilize the \cite{2021arXiv211004576L} calibration, which combines the aforementioned LMC/SMC JAGB zeropoint with a calibration based on MW parallaxes because of the large systematic uncertainties on the Gaia parallaxes used in that study.}
We also adopted the reddening due to the Galactic foreground from the online IRSA Galactic Dust Reddening and Extinction tool\footnote{\url{https://irsa.ipac.caltech.edu/applications/DUST/}} that queries the \cite{1998ApJ...500..525S} full-sky Galactic dust map recalibrated by \cite{2011ApJ...737..103S}.  The extinction value for M33 was determined to be $A_J=0.030$~mag.  Similar to \cite{2017ApJ...845..146H, 2020arXiv200804181J}, we adopted half of the reddening, 0.015 mag, as its systematic uncertainty. 

\subsubsection{Internal Reddening for the JAGB Stars}

Measuring the contribution from reddening internal to M33 and from the dust in the atmospheres of JAGB stars themselves are both challenging. That being said, internal reddening is not expected to be a significant systematic for the JAGB method \textit{in the outer disk of M33}. 
In a study of carbon star distances to M33, \cite{2005A&A...434..657B} quoted negligible internal reddening for the carbon stars because of their large radial distance in the northern outskirts of M33 ($\delta>31^{\circ}$, also where region 3 is located). 
Further indication that internal reddening is not significantly affecting the peak of the JAGB luminosity function in the outer disk of M33 is evidenced in regions 3 and 4, where the JAGB apparent magnitude is roughly constant, and where the statistical uncertainty on the individual measurements is equal to the difference between them (0.01~mag). Therefore, internal reddening is likely not a significant systematic for the JAGB stars \textit{in the outer disk}. Nevertheless, we adopt 0.01~mag as the systematic uncertainty due to internal reddening. 

The circumstellar dust around the JAGB stars also does not likely contribute a significant systematic bias in the JAGB distance scale. In a comparison of TRGB$-$JAGB derived distances in 34 galaxies, \cite{jagbpaper4} found an inter-method scatter of $\pm0.08$~mag. Because circumstellar reddening has a negligible contributions on RGB stars, a significant bias resulting from circumstellar (or internal reddening to the host galaxy, metallicity, star formation history, etc.) would be reflected as a large scatter between the TRGB and JAGB derived distances, and it is not. That being said, we still plan to explore effects of reddening on JAGB stars in greater detail in the next paper in this installment (Lee et al. in prep).

\begin{deluxetable}{ccc}
\tablecaption{JAGB Error Budget (Ground) \label{tab:jagberrors}}
\tablehead{
\colhead{Source of Uncertainty} & 
\colhead{$\sigma_{stat}$} & 
\colhead{$\sigma_{sys}$} \\
\colhead{}&
\colhead{(mag)}& 
\colhead{(mag)}
}
\startdata
Galactic Extinction& \nodata & 0.015\\
Internal Reddening & \nodata &0.01 \\
Scatter in Observed LF (error on the mean) & 0.01 & \nodata \\
Zeropoint & 0.01 & 0.04\\
Choice of $\sigma_s$ & 0.03 & \nodata \\
\hline
\hline
Cumulative Errors & 0.03 & 0.04
\enddata
\end{deluxetable}

Adding the uncertainties in quadrature resulted in a final distance modulus to M33 based on the JAGB method of $\mu_0 = 24.67 \pm0.03$ (stat) $\pm~0.04$ (sys)~mag. 
The error budget for this measurement can be found in Table \ref{tab:jagberrors}.

\subsection{HST}

While the 2MASS $(J-K)$ and $(J-H)$ colors have been shown to be effective in isolating the carbon-rich JAGB stars from their bluer O-rich AGB predecessors, the HST color of F110W$-$F160W (similar to $J-H$) pushes the carbon stars \textit{bluer} than the O-rich stars, effectively making the two populations indistinguishable in color-magnitude space \citep{2012ApJS..198....6D}. Thus, we instead use the HST filter combination F814W-F110W to isolate the JAGB stars. 
At the end of this section, we demonstrate the effectiveness of our adopted color limits by comparing the resulting space-based distance modulus with that from the ground using the conventional $(J-K)$ color. 

In \cite{jagbpaper4}, we provisionally adopted $1.5<(F814W-F110W)<2.5$~mag as the JAGB color limits. 
However, in this paper and moving forward we opted  to more robustly determine the JAGB color limits in the HST photometric system.
To acquire an approximate first sense of where the JAGB stars are in $(F814W-F110W)$ color, we utilized theoretical isochrones to map the JAGB stars from the 2MASS photometric system to the HST photometric system.
\cite{2019MNRAS.485.5666P, 2020MNRAS.498.3283P} recently constrained detailed models of  TP-AGB isochrones by matching them to observations of the LMC and SMC.
We examined these isochrones, which include the JAGB stars in their evolutionary progression, in the different photometric systems: 2MASS's $J$ vs. ($J-K$) and HST's F110W vs. $(F814W-F110W)$ to compare their expected evolutionary tracks.

\begin{figure*}
\centering
\includegraphics[width=\textwidth]{"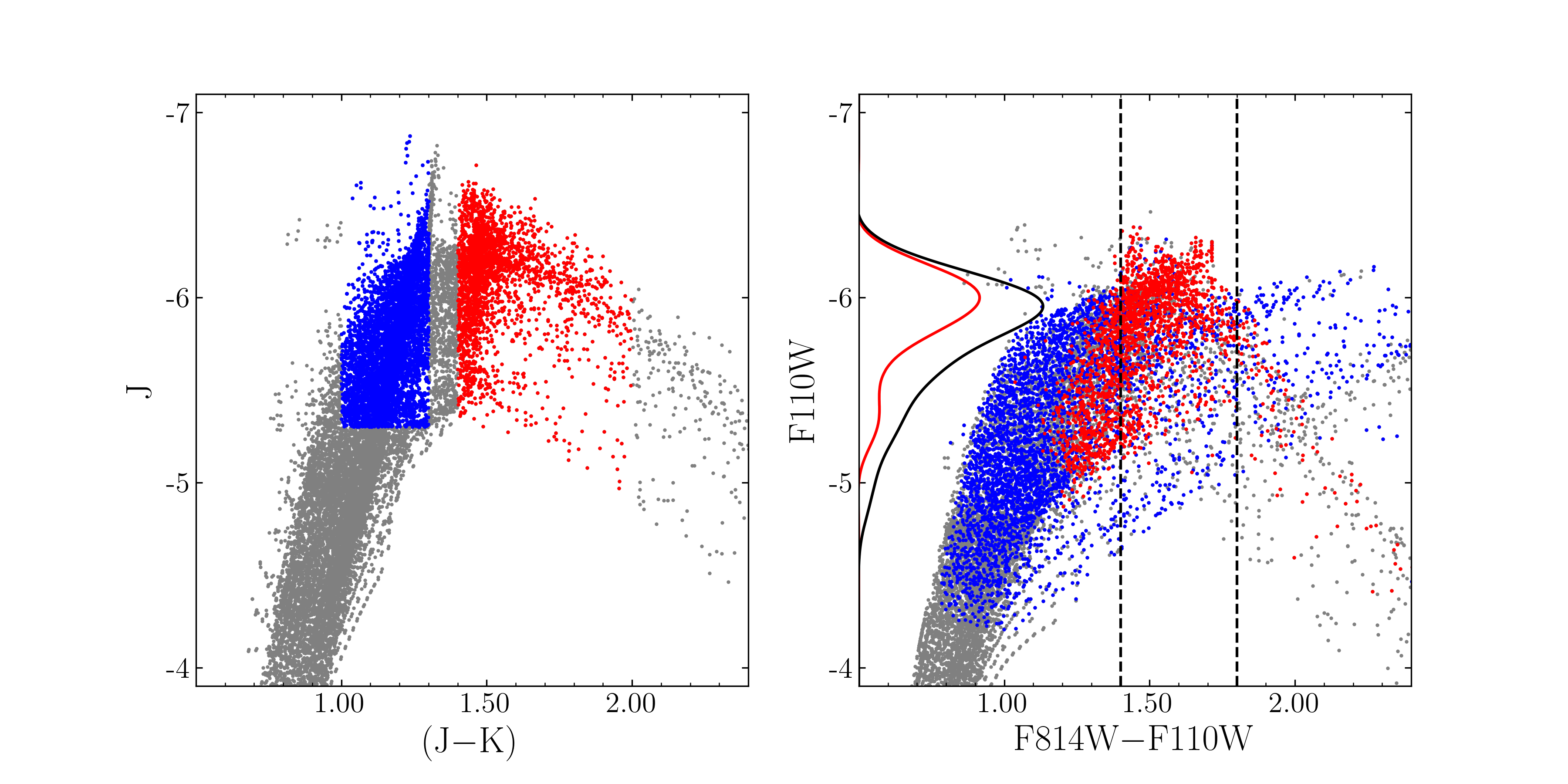"}
\caption{Color magnitude diagrams generated using \texttt{COLIBRI} isochrones for TP-AGB stars. The blue O-rich AGB stars were photometrically selected using a color of $1.0<(J-K)<1.3$~mag and magnitude of $J<-5.3$~mag. The red JAGB stars were selected using a color of $1.4<(J-K)<2.0$~mag. The luminosity functions for all the stars in the color range of $1.4<(F814W-F110W)<1.8$~mag is shown in black. The luminosity function for only the red stars in color range of $1.4<(F814W-F110W)<1.8$~mag is shown in red. } 
\label{fig:fjagb}
\end{figure*}

We generated the theoretical stellar isochrones using the \texttt{PARSEC-COLIBRI} software\footnote{Publicly available at \url{http://stev.oapd.inaf.it/cgi-bin/cmd}} (CMD Version 3.6; \citealt{2012MNRAS.427..127B}; \citealt{2017ApJ...835...77M}) to study the JAGB's expected theoretical behavior, similar to the analysis performed by \cite{2021MNRAS.501..933P}.   
We generated isochrones of stars in the TP-AGB phase with a range of metallicities spanning $-1<[M/H]<0$~dex and ages spanning $1.5-2.5 \times 10^{9}$~years, where carbon star formation is strongly favored \citep{2015ASPC..497..229M, 2017ApJ...835...77M}, shown in Figure \ref{fig:fjagb}. The dots mark points along a given isochrone. The resolution of the thermal pulse cycles, $n_{inTPC}$, was set to 100, where \cite{2017ApJ...835...77M} state that for detailed population studies aimed at accurately reproducing star counts, $n_{inTPC}$ should be increased to $>20$.
We then mapped the JAGB stars within the color and magnitude limits of $1.4<(J-K_s)<2.0$~mag into the F814W and F110W filters, shown in red in Figure \ref{fig:fjagb}. The O-rich AGB stars are also shown in blue, defined by $1.0<(J-K_s)<1.3$~mag and $J<-5.3$~mag. Extreme carbon stars (also known as extreme AGB stars), in the last stage of a typical main-sequence star's lifetime, lie at the reddest end of the CMD at $(J-K)>2.0$, where the majority of them are likely in the `superwind' phase with high mass-loss rates in excess of $\dot M>10^{-6} M_{\odot} ~\rm{yr^{-1}}$ \citep{2017ApJ...835...77M, 2019MNRAS.485.5666P}. 

\begin{figure}
\centering
\includegraphics[width=\columnwidth]{"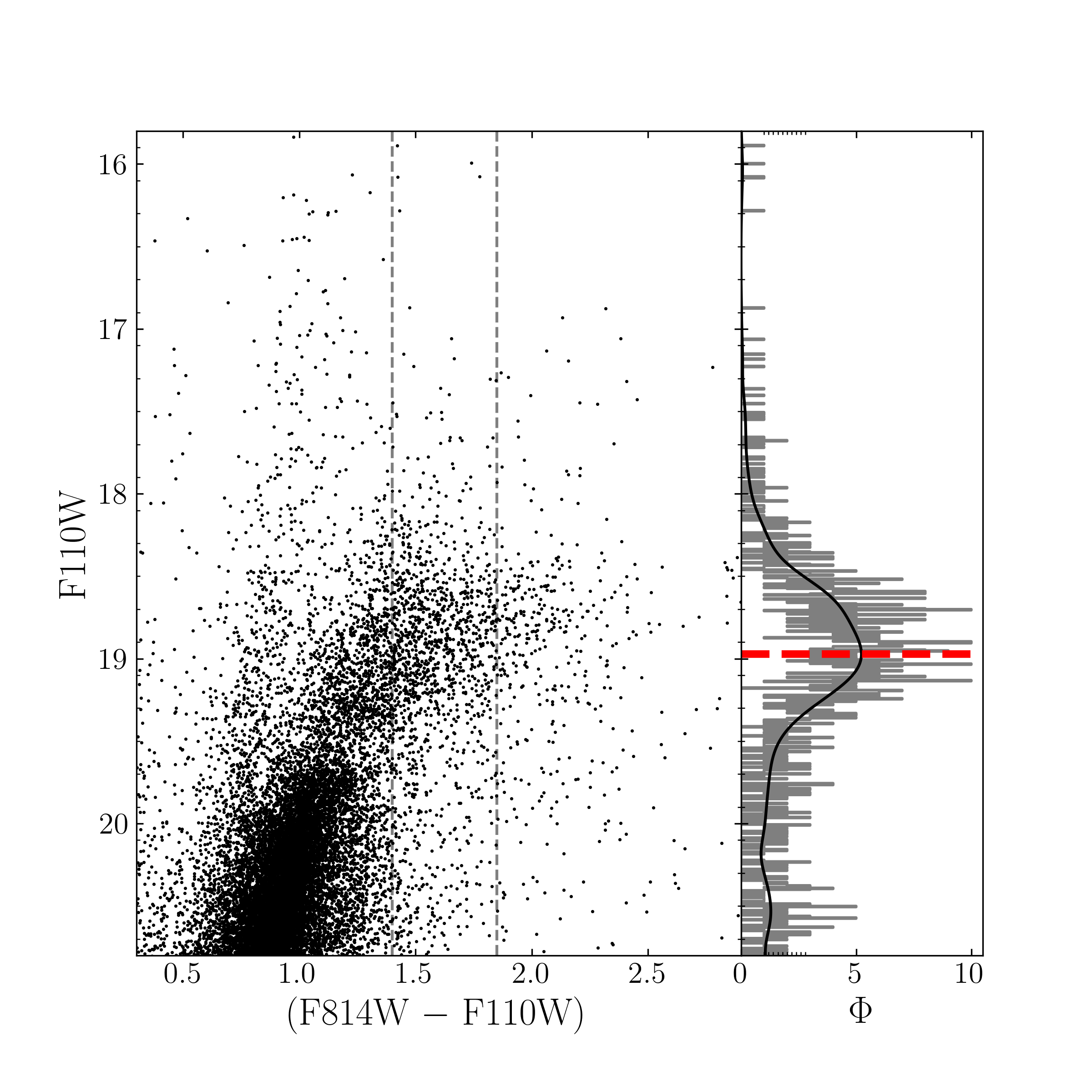"}
\caption{F110W vs. (F814W$-$F110W) color magnitude diagram and smoothed luminosity function in black, overplotted on the binned luminosity function in grey. The JAGB stars can be seen from $1.4<(F814W-F110W)<1.8$~mag. The NIR TRGB can also be seen clearly at around $F110W=19.5$~mag. We measured the JAGB apparent magnitude where the smoothed luminosity function was greatest, marked by the dotted line.}
\label{fig:f10}
\end{figure}

In Figure \ref{fig:f10}, we show the PHATTER F110W vs. F814W-F110W color-magnitude diagram and luminosity function. 
To choose the new HST color limits, upon inspection of the \texttt{COLIBRI} isochrones and PHATTER CMD, we iterated through a blue color cut from  [1.40,1.45,1.50]~mag and a red color cut from [1.65, 1.70, 1.75, 1.80]~mag (i.e. 12  possible combinations). The peak of the 12 possible smoothed JAGB luminosity functions deviated by at most $18.97^{+0.03}_{-0.04}$~mag (using a $\sigma_s=0.20$~mag), so we adopted 0.04~mag as a systematic uncertainty due to our color selection. We adopted the largest range of $1.4<F814W-F110W<1.8$~mag for the JAGB stars' new color limits, and then binned the JAGB star F110W LF at the 0.005~mag level. We then smoothed the binned LF using the GLOESS algorithm as in Section \ref{subsec:ground}. Changing the smoothing parameter from a range of [0.05, 0.20]~mag varied the JAGB apparent magnitude by at most $18.97^{+0.01}_{-0.06}$~mag (using a $\sigma_s=0.20$~mag), so we adopted 0.06~mag as the systematic uncertainty due to the smoothing parameter choice. The peak magnitude of the smoothed LF (using an $\sigma_s=0.20$~mag) was measured to be $18.97\pm0.02$~mag (error on the mean) based on 1,400 JAGB stars. 
As the three PHATTER fields used in this study were located further into the disk than the ground-based JK data used in Section \ref{subsec:ground}; they are mostly situated in region 2, we chose to adopt a slightly larger internal reddening uncertainty of 0.03~mag.
Using the calibration from \cite{jagbpaper4}, $-5.77 \pm0.02$ (stat)~mag, and an extinction value of $A_J \sim A_{F110W}=0.030\pm0.015$, we measured a final JAGB distance modulus in the HST F110W data of $24.71\pm0.06$ (stat) $\pm ~0.05$(sys)~mag, in agreement with our ground JAGB measurement. The error budget for this measurement can be found in Table \ref{tab:jagbhsterrors}. 

We also examined how potential contamination from the blue O-rich stars affect the JAGB peak magnitude in F110W. Using the isochrones from Figure \ref{fig:fjagb}, we compared the peak of the luminosity function for only the red stars (i.e. the stars with $1.4<(J-K)<2.0$~mag) with the peak of the luminosity function for all the stars, in the color range $1.4<(F814W-F110W)<1.8$~mag. The smoothed luminosity functions ($\sigma_s=0.2$~mag) for the two are shown in red and black, respectively. The peak of the luminosity function for all the stars was 0.05~mag fainter, meaning the O-rich stars biased the peak slightly fainter. However, empirically, our ground-based JAGB measurement agrees with our space-based JAGB measurement to within the quoted uncertainties, demonstrating two things: (1) The F110W zeropoint provisionally adopted in \cite{jagbpaper4} appears robust (2) Our newly adopted color limits are also effective at isolating the JAGB stars and for rigorously determining distances in the HST photometric system. These new color limits will be invaluable as we continue to explore and push the application of the JAGB method to farther and farther distances using space-based facilities. We plan to continue to empirically test these new color limits in future papers using HST data.

\begin{deluxetable}{ccc}
\tablecaption{JAGB Error Budget (HST) \label{tab:jagbhsterrors}}
\tablehead{
\colhead{Source of Uncertainty} & 
\colhead{$\sigma_{stat}$} & 
\colhead{$\sigma_{sys}$} \\
\colhead{}&
\colhead{(mag)}& 
\colhead{(mag)}
}
\startdata
Galactic Extinction& \nodata & 0.015\\
Internal Reddening & \nodata &0.03 \\
Scatter in Observed LF (error on the mean) & 0.02 & \nodata \\
Zeropoint & 0.02 & \nodata\\
Choice of $\sigma_s$& 0.06 & \nodata  \\
Color Selection & \nodata & 0.04 \\
\hline
\hline
Cumulative Errors & 0.06 & 0.05
\enddata
\end{deluxetable}

\section{Tip of the Red Giant Branch}\label{sec:trgb}

The TRGB has emerged in the past several decades as a highly accurate and precise local distance indicator.
The TRGB has a well-understood theoretical basis, marking the core helium-flash for all low-mass red giant stars \citep{1997MNRAS.289..406S}.  At this point in the red giant star's lifetime, the temperature of its electron-degenerate helium core has reached $\sim 10^8$~K from hydrogen-shell fusion, lifting the electron degeneracy and triggering the onset of core-helium fusing through the triple-$\alpha$ process. This results in an abrupt decrease in the star's luminosity as the star settles onto the horizontal branch. Empirically, this transition in a red giant's evolution reveals itself as a sharp discontinuity in a galaxy's luminosity function, from which the TRGB can be accurately and robustly determined \citep{1993ApJ...417..553L, 2019ApJ...882...34F}. In the I band, the TRGB is remarkably constant over a range of metallicities and ages \citep{2020ApJ...891...57F}, providing an excellent standard candle, capable of measuring highly accurate distances out to about $\sim30$~Mpc \citep{2017ApJ...836...74J}. 

Recent studies have also shown the NIR TRGB, while color-dependent and upward-sloping, has advantages to the I-band TRGB (e.g. \citealt{2012ApJS..198....6D, 2014AJ....148....7W, 2018ApJ...858...12H, 2018ApJ...858...11M, 2020ApJ...898...57D}). Red giant stars are brighter in the NIR wavelengths than in the optical, and therefore can be used to probe farther distances. Furthermore, effects of dust extinction and reddening are significantly decreased in the NIR. Using JWST, the infrared TRGB can be extended out to $>40$~Mpc (i.e. a volume five times greater than currently possible with HST;  \citealt{2021arXiv210615656F}). Continuing to explore the NIR TRGB will be increasingly relevant as JWST's launch nears. In this paper, we measure the distance modulus to M33 based on both the I-band and NIR TRGB.

\subsubsection{JHK-band TRGB}\label{subsubsec:nir}

\begin{figure*}
\centering
\includegraphics[width=\textwidth]{"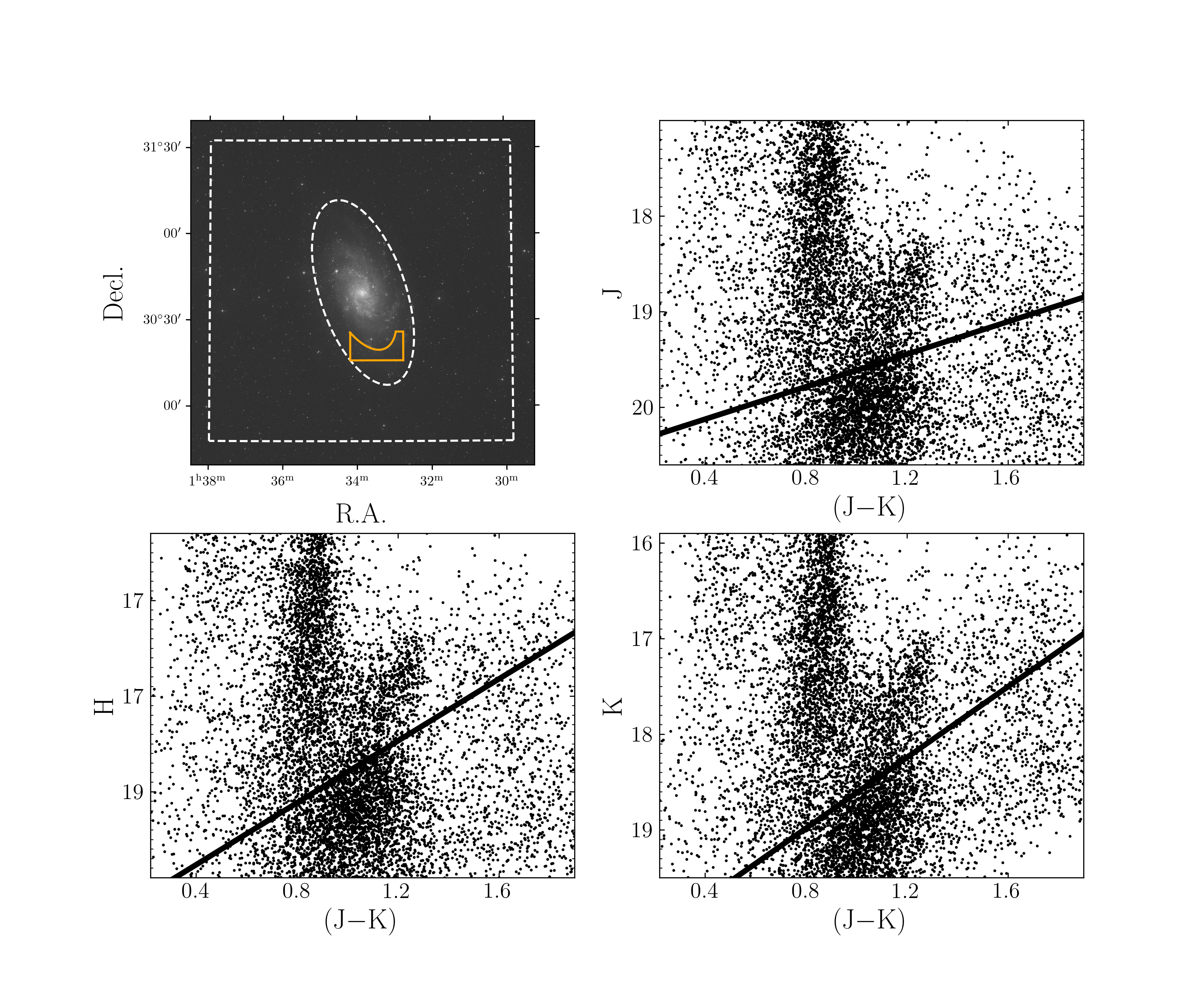"}
\caption{NIR TRGB color magnitude diagrams for stars in the outer region of M33. Only 1 star in 3 is plotted so the TRGB discontinuity can be easily seen. The solid black line in each plot shows the adopted TRGB. The spatial selection for the NIR TRGB stars is shown in the upper left hand panel by the region between the dashed contours. The orange area marks where our I-band TRGB measurement was performed. 
}
\label{fig:f3}
\end{figure*}

It has been well established that TRGB measurements should avoid being made in the disks of galaxies where contamination from younger stellar populations and dust can significantly bias the tip detection \citep{2020arXiv200804181J, 2021arXiv210613337H}.
Therefore, we first performed a spatial cut on our JHK data, only selecting stars in the lower-dust and sparser outer regions of M33, shown by the area inside the white dotted lines in Figure \ref{fig:f3}. Also in Figure \ref{fig:f3}, we show color magnitude diagrams of the M33 TRGB stars. The upward-sloping TRGB is marked by the black lines, determined by RGB stars with the color $0.7<(J-K)<1.3$~mag. The TRGB was determined in the following way, consistent with \cite{2018ApJ...858...11M, 2020arXiv201209701C}.\footnote{We chose not to use the tracer star method as employed in \cite{2021ApJ...907..112L, 2020ApJ...891...57F}, which maps tip stars in the I band into near-infrared wavelengths, as our VI dataset only covered a small section of M33 (see Figure \ref{fig:f3}).}
Using the J band as a first illustrative example, we first visually made a first approximation of the zeropoint of the TRGB locus, using predetermined TRGB slopes from \cite{2020ApJ...891...57F}. The data were then ``rectified'', i.e. transformed into the $T[J, (J-K)]$ plane as first introduced and implemented in \cite{2009ApJ...690..389M}, so that the TRGB appeared flat as a function of color. The T-band luminosity function was then created by first binning the color-selected T-band magnitude of the RGB stars using bins of 0.01~mag. The binned LF was then smoothed using the GLOESS algorithm with a smoothing parameter of $\sigma_s=0.10$~mag. The tip was then detected using a Sobel edge detection filter in the same manner as in the I-band TRGB, which is further described in Section \ref{subsec:itrgb}.

The independent tip detections for each of the rectified $JH$ luminosity functions gave the following apparent magnitude relations for the TRGB in M33\footnote{We omit K because $m_J^{TRGB}$ and $m_K^{TRGB}$ are not independent by requirement, see \cite{2020AJ....160..170M}.}:

\begin{figure*}
\gridline{\fig{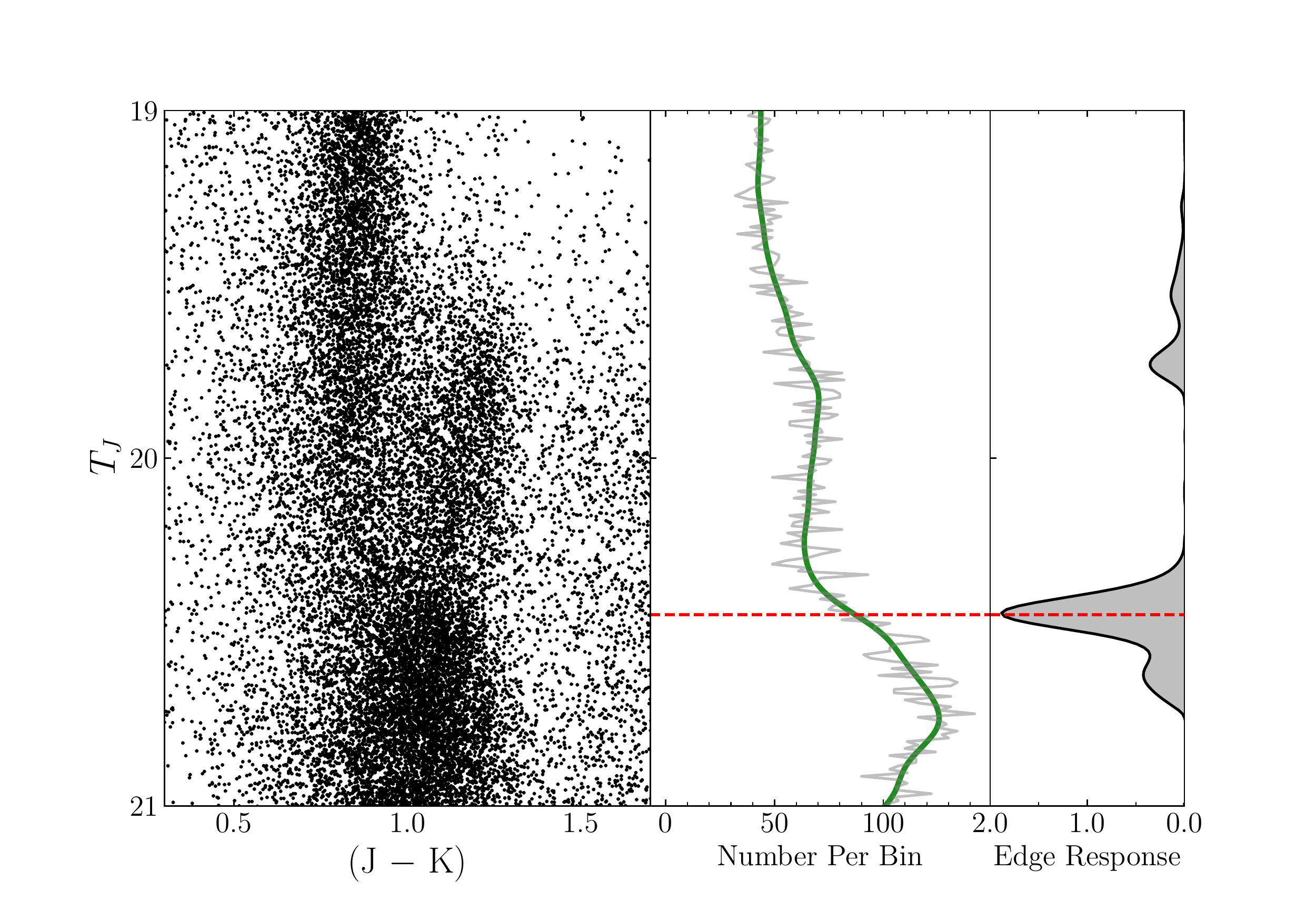}{0.5\textwidth}{}
          \fig{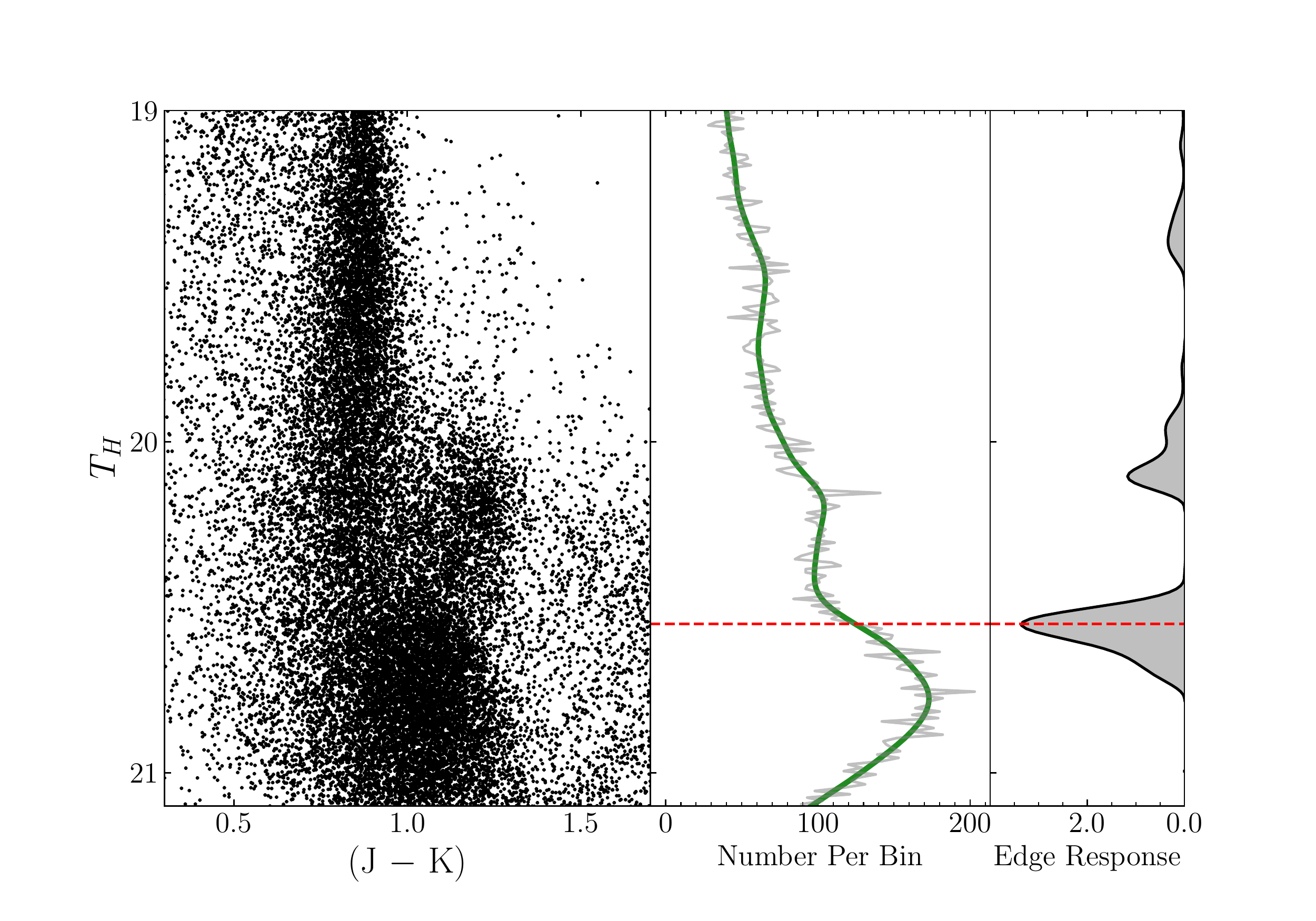}{0.5\textwidth}{}}
\caption{Rectified J and H vs. ($J-K$) color-magnitude diagram for stars in the outer regions of M33. Star magnitudes were rectified based on the slopes from \cite{2020ApJ...891...57F}, so that the TRGB discontinuity is flat. The TRGB is then determined using an edge detector as is normally done with the I-band TRGB.
\label{fig:okey}}
\end{figure*}

\begin{equation}
m_J^{TRGB}=19.62-0.85 \times [(J-K)_o - 1.00]
\end{equation}
\begin{equation}
m_H^{TRGB}=18.79-1.62\times[(J-K)_o - 1.00]
\end{equation}

The average photometric uncertainty of the  stars $J=19.7\pm0.2$~mag was 0.08~mag and 0.06~mag for J and H respectively, which we adopted as systematic errors on the final NIR TRGB measurement. The width of the filter response at the measured T-band, i.e. the statistical uncertainty in measuring the peak, was approximately $\pm0.04$~mag for both J and H.
We also applied foreground extinction corrections: $A_J=0.030\pm0.015$~mag and $A_H=0.019\pm0.0095$~mag.

Several studies have quoted negligible internal reddening for the TRGB stars in the outer regions of M33, such as \cite{tmp} and \cite{2009ApJ...704.1120U}, with the former's assumption based on detailed FIR dust maps from \cite{2003A&A...407..137H}. Later in Section \ref{sec:ceph}, we measure an internal extinction for the Cepheids in the outer disk of $E(B-V)=0.12$~mag, which translates into an $A_J=0.10$~mag. As our TRGB measurement is even farther out into the halo and RGB stars are expected to be significantly less affected by reddening than Cepheids, we also assumed negligible  contribution from internal reddening for the TRGB stars in the halo of M33. 

We then determined distance moduli based on the absolute calibrations from \cite{2018ApJ...858...12H}, which we repeat below. The errors on these zeropoints were quoted as $\pm0.01$ (stat) and $\pm0.06$ (sys).
\begin{equation}
M_J^{TRGB}=-5.14-0.85 \times [(J-K)_o - 1.00]
\end{equation}
\begin{equation}
M_H^{TRGB}=-5.94-1.62\times[(J-K)_o - 1.00]
\end{equation}

Our final measured NIR TRGB distance moduli were then: 
$\mu_0 (TRGB_J)=24.73\pm 0.04$ (stat) $\pm~ 0.10$ (sys) and 
$\mu_0 (TRGB_H)=24.71\pm 0.04$ (stat) $\pm~ 0.09$ (sys), for a final combined NIR TRGB measurement of $\mu_0 (TRGB_{NIR})=24.72\pm 0.04$ (stat) $\pm~ 0.10$ (sys).

\begin{deluxetable}{c|cc|cc}
\tablenum{2}
\tablecaption{NIR TRGB Error Budget \label{tab:nirtrgberrors}}
\tablehead{
\multicolumn{1}{c}{ } &
\multicolumn{2}{c}{$TRGB_J$} &
\multicolumn{2}{c}{$TRGB_H$}  \\
\colhead{Source of Uncertainty} & 
\colhead{$\sigma_{stat}$} & 
\colhead{$\sigma_{sys}$} &
\colhead{$\sigma_{stat}$} & 
\colhead{$\sigma_{sys}$}  \\
\colhead{} & 
\colhead{(mag)} & 
\colhead{(mag)}& 
\colhead{(mag)}& 
\colhead{(mag)}
}
\startdata
Extinction& \nodata  & 0.015 & \nodata & 0.0095 \\
Zeropoint & 0.01 & 0.06 & 0.01 & 0.06\\
Photometric Uncertainties & \nodata  & 0.08 & \nodata & 0.06 \\
Width of Edge Response & 0.04 & \nodata & 0.04 & \nodata \\
\hline
\hline
Cumulative Errors & 0.04 & 0.10 & 0.04 & 0.09
\enddata
\end{deluxetable}

\subsection{I-band TRGB}\label{subsec:itrgb}

\begin{figure*}
\centering
\includegraphics[width=.9\textwidth]{"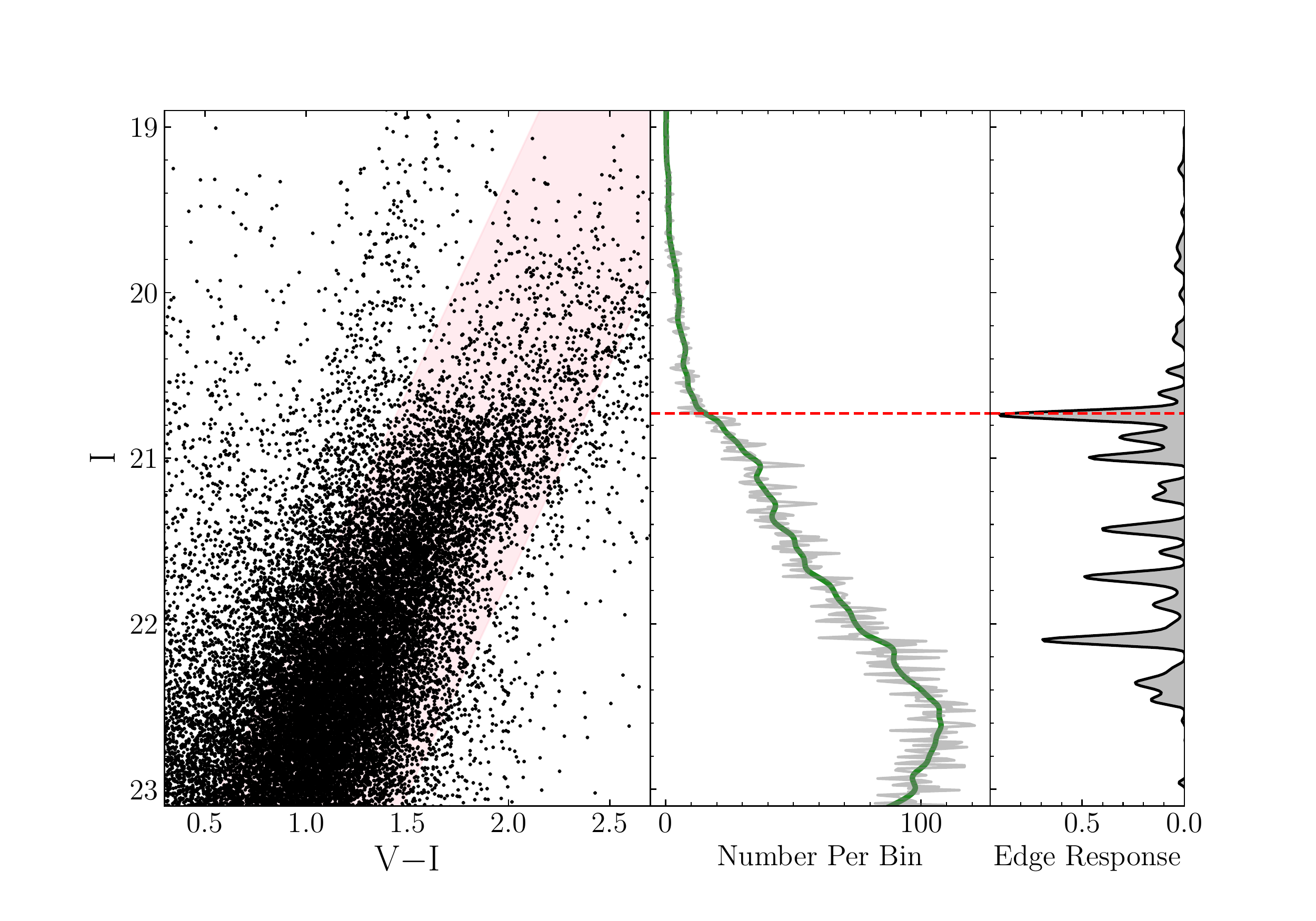"}
\caption{(Left) I vs. (V$-$I) color magnitude diagram. The color selection used to measure the TRGB is marked by the pink shaded region. (Middle) Binned I-band luminosity function in grey, with the GLOESS smoothed LF over-plotted in green. (Right) Sobel Filter Edge Response function: its greatest value marks the largest discontinuity in the smoothed I-band luminosity function and thus empirically marks the TRGB, indicated by the dotted red line. The smaller, less prominent peak at $I\sim22$~mag visually lies firmly with the RGB itself and does not correspond to any known features; thus, we associate only the brighter (and more prominent) peak with the TRGB.}
\label{fig:f2}
\end{figure*}

To detect the I-band TRGB, we followed the CCHP approach first described in  \cite{2017ApJ...845..146H}. First, we spatially excluded data in the spiral arm of M33 by only using the data in the southern-most part of our sample, shown in Figure \ref{fig:f3} by the orange panel. This spatial selection ensured we were avoiding crowding and dust effects in the star-forming spiral arm region of that dataset, also where the Cepheids reside. Then, we color-selected the metal-poor blue TRGB stars; the selection box is shown by the pink shaded area in the left panel of Figure \ref{fig:f2}. 
We then finely binned their I-band luminosity function using bins of 0.01~mag. The binned luminosity function was then smoothed using a GLOESS algorithm to reduce Poisson noise peaks, using a smoothing parameter of $\sigma_s=0.075$~mag. Next, the smoothed luminosity function was convolved with a Sobel kernel [-1,0,+1] (first derivative) edge detection filter, resulting in an edge response function which measured the gradient of the smoothed luminosity function. The greatest value of the edge response function is the point of the largest discontinuity in the luminosity function and marks the TRGB.

In Figure \ref{fig:f2}, we show the $I$ vs. $(V-I)$ color magnitude diagram, luminosity function, and edge response function, with the measured TRGB marked by a red dashed line. 
The width of the edge response function is extremely clear and narrow.
Thus, we conservatively adopted a 0.01~mag statistical uncertainty and a 0.01~mag systematic uncertainty on our TRGB measurement based on previous artificial star experiments performed by the CCHP (e.g. \citealt{2017ApJ...845..146H, 2020arXiv200804181J}). 
The median photometric uncertainty $\pm0.10$~mag above and below the TRGB was $\pm0.023$~mag, which we also adopted as a systematic uncertainty.
We also added $\pm0.019$~mag as an additional systematic uncertainty on the photometric zeropoint used to calibrate the data, following Table 3 of \cite{2009MNRAS.396.1287S}. All of the tabulated errors were added in quadrature to obtain the final systematic and statistical errors. 
The resulting I-band TRGB magnitude in M33 was measured to be $m^{TRGB}_I=20.73 \pm 0.01$ (stat) $\pm ~0.03$ (sys)~mag.

To measure the distance modulus, we utilized the TRGB calibration from \cite{2020ApJ...891...57F}: $M_I^{TRGB} = -4.05 \pm 0.02$ (stat) $ \pm$ 0.04 (sys), and a galactic extinction value of $A_I=0.063\pm0.032$ (sys)~mag. Because the I-band TRGB field is closer into the disk than the NIR TRGB, we conservatively added an additional systematic uncertainty for internal reddening of 0.03~mag to the I-band TRGB error budget.
In conclusion, our final measured apparent distance modulus to M33 based on the I-band TRGB was determined to be $\mu_0 (TRGB_I)=24.72 \pm 0.02$ (stat) $\pm~0.07$ (sys)~mag. The error budget is tabulated in Table \ref{tab:trgbierrors}.

\begin{deluxetable}{ccc}
\tablecaption{I-band TRGB Error Budget \label{tab:trgbierrors}}
\tablehead{
\colhead{Source of Uncertainty} & 
\colhead{$\sigma_{stat}$} & 
\colhead{$\sigma_{sys}$} \\
\colhead{}&
\colhead{(mag)}& 
\colhead{(mag)}
}
\startdata
Galactic Extinction& \nodata & 0.032\\
Internal Reddening & \nodata & 0.03 \\
Zeropoint & 0.02 & 0.04\\
Photometric Uncertainty & \nodata & 0.023\\
Photometric Zeropoint & \nodata & 0.019\\
Artificial Star tests & \nodata & 0.01\\
Width of Edge response function & 0.01 & \nodata \\
\hline
\hline
Cumulative Errors & 0.02 & 0.07
\enddata
\end{deluxetable}

\section{Cepheid Period-Luminosity Relations}\label{sec:ceph}
In this section, we determine a distance to M33 based on its Cepheid period-luminosity relation. Cepheids are young, yellow supergiants found in the disks of star-forming, late-type spiral galaxies.
They have a well-defined relation between their period, luminosity, and color, and for many decades have served as the gold standard of local distance measurements (e.g., the HST Key Project: \citealt{2001ApJ...553...47F} and \citealt{1999ApJ...522..802S}). Furthermore, Cepheids have many well-established strengths that make them powerful distance indicators: (a) Their high intrinsic brightness ($-2<M_V<-6$) have allowed distances up to $\sim40$~Mpc to be probed. (b) Due to their variability, Cepheids are easily identified and classified, and their periods are sufficiently stable over a human lifetime. (c) In the infrared, their PL relations have small intrinsic dispersion. 

However, several potential systematic effects unique to the Leavitt law become serious challenges at farther distances, including photometric errors due to crowding/blending in the inner disks of galaxies (especially at infrared wavelengths), dust contamination in the disk, and a potential additional uncertainty in a metallicity dependence of the Leavitt law \citep[e.g.,][]{2020arXiv200710716E, 2021arXiv210615656F, 2021arXiv211008860R}. 
Many of these concerns remain unresolved and may be sources of potential systematic uncertainty for the Cepheid distance scale. We measured a Leavitt law distance to M33 to use for comparison with the TRGB and JAGB, and to emphasize empirically that for relatively nearby galaxies like M33, the TRGB, Leavitt Law, and JAGB method agree very well. It is only for farther distances that the TRGB and Leavitt law begin to diverge, and where the JAGB will help to act as an arbitrator.

\subsection{The Cepheid Sample}\label{subsec:ceph}

To measure a Cepheid distance modulus to M33, we targeted Cepheids in the outer regions of M33, where effects of crowding and blending are smaller.  We identified two catalogs of outer region Cepheids in M33. The first, which comes from the outer field sample of  \cite{2009MNRAS.396.1287S}, contained BVI photometry and periods for 40 Cepheids ranging from $20.3<V<23.0$~mag in the southern spiral arm of M33. The second, comes from \cite{2011ApJS..193...26P} and contained BVI photometry for 564 Cepheids ranging from $17.8<V<22.8$~mag distributed across M33's entire disk. 
There were 4 Cepheids in common between the two catalogs, for which we averaged the periods and VI magnitudes.  We chose to exclude the B band in this analysis because of its potential strong metallicity dependence \citep{1991PASP..103..933M}.

We restricted our catalog to Cepheids in the outer regions of M33 (see Figure \ref{fig:f1}, about 5~kpc away from the center of M33 on the major axis)\footnote{See Section \ref{subsec:optphot} for more details.}. We also chose to only use Cepheids with periods greater than 10 days, given the contamination by overtone pulsators and the possibility that the Cepheid P-L relation may exhibit nonlinearities for Cepheids with periods less than 10 days \citep{2004A&A...424...43S, 2009ApJ...693..691N}. In the end, 60 Cepheids remained. As the width of the P-L relation drastically decreases in going from the optical to the near infrared, random-phase observations of Cepheids in the near-infrared are comparable in accuracy to complete time-averaged magnitudes in the blue \citep{1991PASP..103..933M}. Thus, we were able to supplement the BVI observations with our $JHK$ photometry. In the end, we were able to locate 53 of the Cepheids in our $JHK$ photometry using a matching radius of $1\arcsec$. 

\subsection{Cepheid P-L Relations and Reddening Curve Fit}\label{subsec:cephred}

                \begin{figure}
\centering
\includegraphics[width=\columnwidth]{"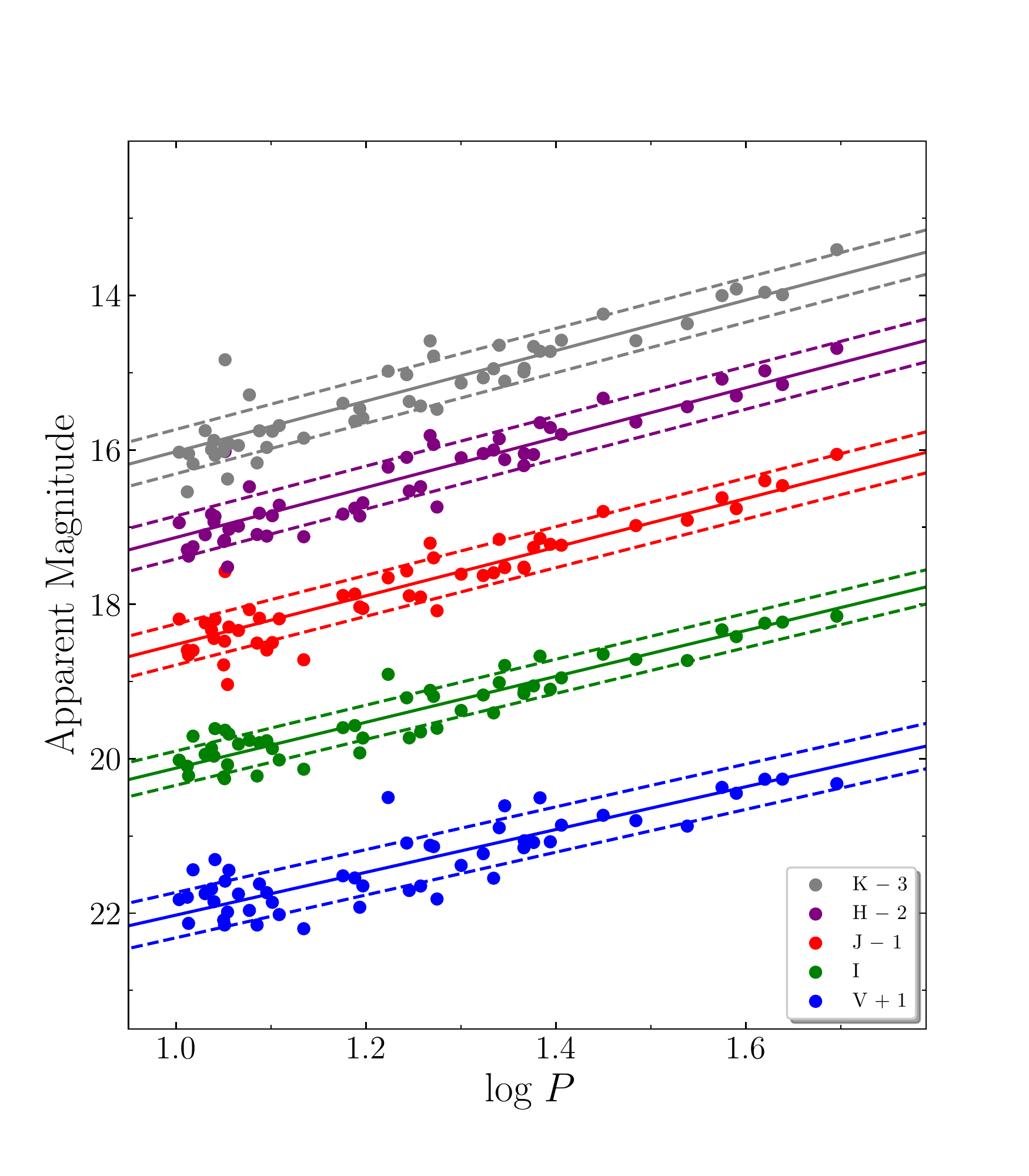"}
\caption{Period-luminosity relations for Cepheids in the VIJHK bands in the outer halo of M33. Slopes were fixed and taken from the Milky Way P-L relations in \citetalias{2012ApJ...759..146M}. These Cepheids have periods ranging from 10 to 50 days. 1 $\sigma$ boundaries are shown as dotted lines.
\label{fig:f6}}
\end{figure}

\begin{figure*}
\centering
\includegraphics[width=\textwidth]{"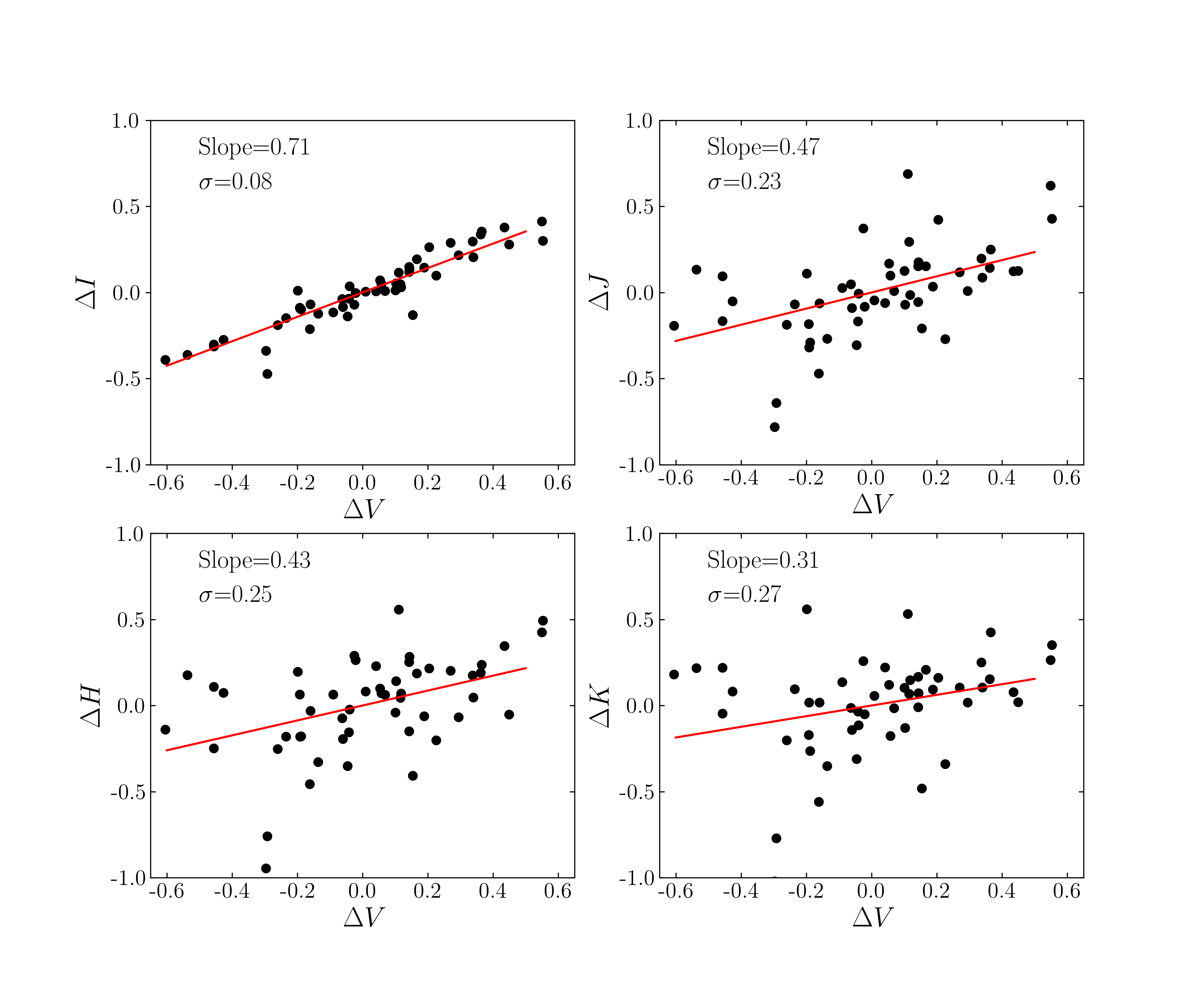"}
\caption{Magnitude residuals from the P-L relations for IJHK vs V. The correlated scatter shows the influence of the Cepheids' intrinsic positions in the instability strip. The increased scatter in JHK compared to I indicate that the widths of their P-L relations are also likely being influenced by random photometric uncertainties.
\label{fig:resid}}
\end{figure*}

VIJHK P-L relations for Cepheids in M33 are shown in Figure \ref{fig:f6}. The P-L fits were determined using fixed slopes from \citealt{2012ApJ...759..146M} (hereafter \citetalias{2012ApJ...759..146M}).  Using the \citetalias{2012ApJ...759..146M} intercepts, we determined wavelength-dependent apparent distance moduli, plotted in Figure \ref{fig:f6}. The correlated magnitude residuals are also plotted in Figure \ref{fig:resid}, demonstrating that the width of the P-L relations in JHK are being more influenced by individual photometric errors than in V and I. This was to be expected as the VI photometry was time-averaged over many epochs, and the JHK photometry was not.

\begin{figure}
\centering
\includegraphics[width=\columnwidth]{"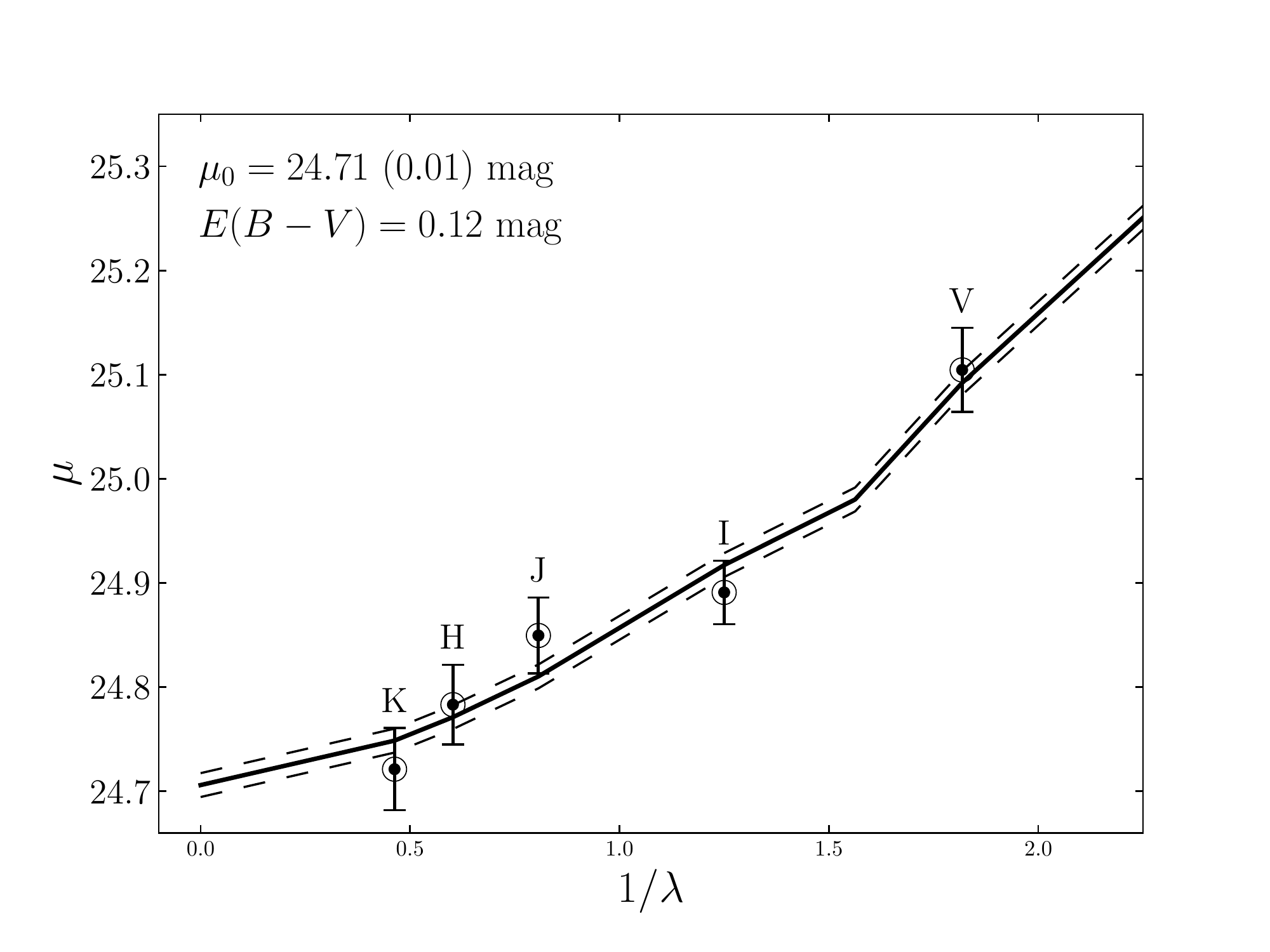"}
\caption{The extinction curve fit to the wavelength-dependent distance moduli derived from M33 outer disk Cepheids. The solid line shows the best fit using a chi-squared grid minimizer, and the dashed lines give the 1-$\sigma$ error. Error bars on the individual points give their standard error on the mean from the fit on their P-L relations. 
\label{fig:f7}}
\end{figure}

We then fit the \cite{1989ApJ...345..245C} extinction law to the wavelength-dependent distance moduli to simultaneously solve for the $E(B-V)$ color excess and true distance modulus: $A(\lambda)/A_V = a(x) + b(x)/R_v$, with $R_v=3.1$, and where $a=0.574x^{1.61}$, $b=-0.527x^{1.61}$, and $x=1/\lambda$. For $\lambda$, we used 
$VIJHK$ = 0.55, 0.80, 1.24, 1.66, and 2.16 $\mu$m, following
\citetalias{2012ApJ...759..146M}.
The best-fit true distance modulus was measured to be $\mu_0 = 24.71\pm0.01$ (stat)~mag with a color excess of $E(B-V)=0.12$~mag, shown in Figure \ref{fig:f7}. We note that sigma-clipping the P-L relations, specifically the two Cepheids with residuals of $\Delta H < -0.6$ (see Figure \ref{fig:resid}) resulted in a fainter distance modulus of $24.74$~mag.

\subsection{Bootstrap Error Analysis}
As shown in Figure \ref{fig:f7}, the scatter about the extinction curve fit, measured to be 0.01~mag, was appreciably smaller than the errors on the individual wavelength-dependent distance moduli, which ranged between 0.04 and 0.05~mag. This indicated either that the errors on the distance moduli were overestimated, or the scatter on the fit was underestimated. To diagnose this problem, we performed an additional bootstrapping error analysis to ascertain an accurate statistical error on the measured true distance modulus and reddening. 

We used random sampling with replacement to generate 1,000 samples of 53 Cepheids from the original Cepheid dataset.
Then, for a given sample, we refit the P-L relations and then re-measured the true distance modulus and reddening.  The mean distance modulus was measured to be $\mu_0 = 24.71\pm0.04$~mag in agreement with the actual measured distance modulus in Section \ref{subsec:ceph}; albeit its standard deviation was more comparable with the errors on the individual wavelength-dependent distance moduli. The total color excess was measured to be $E(B-V)=0.12\pm0.01$~mag. 
We adopted the 1 $\sigma$ uncertainties determined by our bootstrapping analyses as the uncertainties on the distance modulus $\mu_0$ and reddening $E(B-V)$. We also adopted the uncertainty on the reddening as the systematic uncertainty on our true distance modulus. Thus, our final adopted distance modulus derived from the M33 Cepheids was measured to be $\mu_0=24.71\pm~0.04$ (stat) $\pm0.01$ (sys) mag.

\section{Independent Distance Comparisons}\label{sec:compre}
\begin{figure*}
\centering
\includegraphics[width=.6\textwidth]{"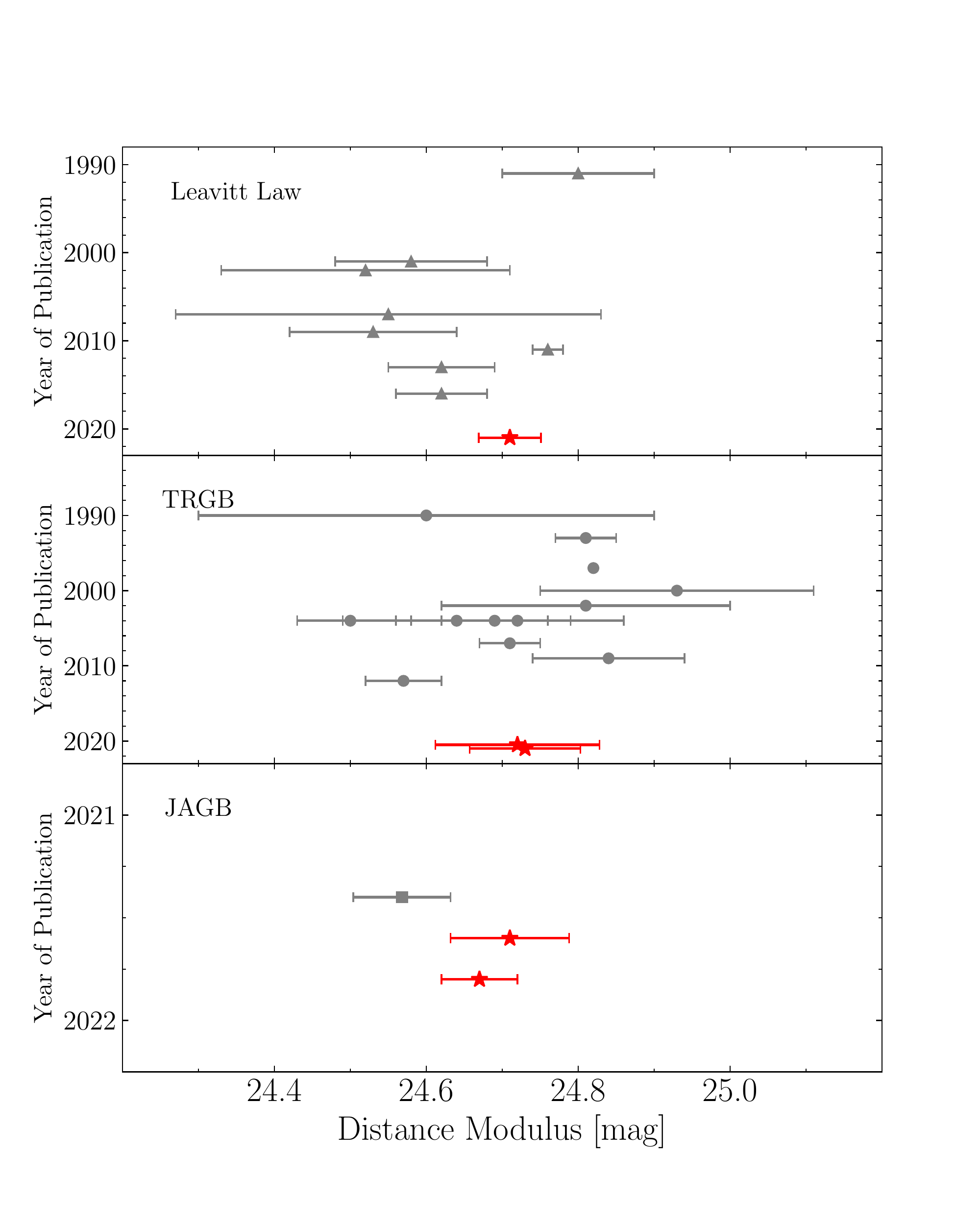"}
\caption{Comparison of literature distances to M33 using the Leavitt Law (top panel), I-band TRGB (middle panel), and JAGB method (bottom panel). In all three panels, the measurements from this study are shown as red stars. For the JAGB, both our space-based (top star) and ground-based (bottom star) measurements are shown. 
}
\label{fig:f11}
\end{figure*}

In this section, we compare our distance measurements to those compiled from the literature. We show a visual compilation of previous Leavitt Law, TRGB, and JAGB measurements in Figure \ref{fig:f11} and tabulate them in Table \ref{tab:compre}.

\begin{deluxetable*}{lclcc}\label{tab:compre}
\tablenum{3}
\tablecaption{Leavitt Law, TRGB, and JAGB Distance Moduli to M33\label{tab:history}}
\tablewidth{1pt}
\tablehead{
\colhead{Study} & 
\colhead{Method} & 
\colhead{$\mu_{0}$ [mag]} & 
\colhead{Notes} 
}
\startdata
\cite{1991ApJ...372..455F} & Leavitt Law & $24.64\pm0.09$\\
\cite{1991MNRAS.250..438M} & Leavitt Law & $24.80 \pm 0.10$\\
\cite{2001ApJ...553...47F} & Leavitt Law & $24.58\pm0.10$\\
\cite{2002ApJ...565..959L} & Leavitt Law & $24.52\pm0.19$\\
\cite{2007ApJ...671.1640A} & Leavitt Law & $24.55\pm 0.28$\\
\cite{2009MNRAS.396.1287S} & Leavitt Law & $24.53 \pm 0.11$\\
\cite{2011ApJS..193...26P} & Leavitt Law & $24.76 \pm 0.02$ & Reddening-free Wesenheit relation ($W_{VI}$) for their cleaned sample \\
\cite{2013ApJ...773...69G} & Leavitt Law & $24.62\pm0.07$\\
\cite{2016AJ....151...88B} & Leavitt Law &  $24.62\pm0.06$\\
This paper & Leavitt Law & $\bf 24.71 \pm 0.04$ & \\
\hline
\cite{1990AJ.....99..149W} & TRGB & $24.60\pm 0.30$ & Central disk \\
\cite{1993ApJ...417..553L} & TRGB & $24.81\pm0.04$\\
\cite{1997MNRAS.289..406S} & TRGB & 24.82 & No uncertainties given \\
\cite{2000ApJ...529..745F} & TRGB & $24.93\pm 0.18$\\
\cite{2002AJ....123..244K} & TRGB & $24.81^{+0.19}_{-0.07}$  & Disk field\\
\cite{2004MNRAS.350..243M} & TRGB & $24.50\pm 0.06$\\
\cite{2004AJ....128..224T} & TRGB & $24.69\pm 0.07$ & Outer disk\\
\cite{2004AJ....128..237B} & TRGB & $24.72 \pm0.14$ & Outer halo \\
\cite{tmp} & TRGB & $24.64\pm0.15$ & Outer halo\\
\cite{2007ApJ...661..815R} & TRGB & $24.71\pm0.04$ & Disk Field\\
\cite{2009ApJ...704.1120U}  &TRGB &  $24.84\pm0.10$ & Halo \\
\cite{2012ApJ...758...11C} & TRGB & $24.57\pm0.05$\\
This paper & TRGB & $\bf 24.72 \pm 0.07$& I-band \\
This paper & TRGB & $\bf 24.72 \pm 0.11$& $JHK$-band \\
\hline
\cite{2021arXiv210502120Z} & JAGB & $24.57 \pm 0.06$ & Inner disk\\
This paper & JAGB & $\bf 24.67 \pm 0.05$ & Ground-based, outer disk \\
This paper & JAGB & $\bf 24.71 \pm 0.08$ & Space-based, outer disk \\
\enddata
\end{deluxetable*}

Since 1990, there have been 10 Leavitt Law distances to M33 published. We found their average to be $\mu_0=24.63\pm0.04$~mag, which is about 1.4~$\sigma$ away from our measurement. 

There has only been one JAGB distance modulus measured to M33 thus far by \cite{2021arXiv210502120Z}. They used \cite{2013ApJ...773...69G} inner disk near-infrared data and measured $\mu_0 = 24.57\pm0.06$~mag, in $\sim 1.3\sigma$ agreement with our ground-based measurement. We note that their CMD (their Figure 6) looks visually similar to our CMDs in regions 1 and 2, with less of a clearly-defined peak than in regions 3 and 4. Moving farther out into the disk may mitigate some of the disagreement between the \cite{2021arXiv210502120Z} measurement and ours. 

For the I-band TRGB (there have been no NIR TRGB distances yet measured to M33), we found 12 distances in the literature, resulting in an average distance modulus of $24.72\pm0.04$~mag, in excellent agreement with both our NIR and I-band TRGB measurement.

We also note that the mean distances from the literature to M33 derived from the Leavitt law and TRGB disagree at the 1.6-$\sigma$ level, with the Leavitt law distances being 0.09~mag brighter on average (corresponding to a non-negligible distance of 35.7~kpc). The reason(s) for this difference are not known at this time.

For context, we also provide the average distances determined from other distance indicators:
\begin{itemize}

    \item \cite{2009ApJ...704.1120U} used the flux-weighted gravity-luminosity relationship (FGLR) to measure a distance modulus of $24.93\pm0.11$~mag. 
    
    \item Modeling observed CMDs, \cite{2011MNRAS.410..504B} measured a distance modulus of $24.69\pm0.09$~mag. 
    
    \item \cite{2006ApJ...652..313B} presented a distance modulus based on detached eclipsing binaries in M33 to be $24.92\pm0.12$~mag.
    
    \item The average estimate from planetary nebula luminosity functions (PNLFs) yields 
    $24.74_{-0.14}^{+0.13}$~mag
    \citep{2004ApJ...614..167C, 2000A&A...355..713M}.
    
    \item \cite{2002AJ....123..244K} measured a distance using the red clump (RC) of $24.80\pm 0.06$~mag. 
    
    \item By monitoring the proper motions of water masers in M33, \cite{2005Sci...307.1440B} measured a geometric distance modulus of $24.31_{-0.18}^{+0.51}$~mag.

    \item And finally, averaging the five RR Lyrae distance moduli \citep{2000AJ....120.2437S,2006AJ....132.1361S, 2010ApJ...724..799Y,2011AJ....142..198P, 2013MNRAS.435.3206D} to M33 yields $24.62\pm0.06$~mag. 
\end{itemize}

Given the uncertainties, our measured distance moduli agree at the 1 $\sigma$ level with estimates from the CMD modeling, PNLFs, and water masers, and disagree with estimates from the FGLR, DEBs, and red clump. The distance modulus derived from RR Lyrae agrees at the 1 $\sigma$ level with both our JAGB method distance moduli and at the 2 $\sigma$ level with the TRGB and Leavitt law.

\section{Summary and Conclusions}\label{sec:sum}
In this paper, we have determined the distance modulus to M33 using three independent methods: the TRGB, JAGB method, and Leavitt law. 
All of the distance moduli agree to within 2\%, providing further evidence in addition to \cite{2021ApJ...907..112L}, that the JAGB method is as precise and accurate as the TRGB and Leavitt law distance indicators. 
We found the four measured distance moduli to be $\mu_0 ~(Cepheids) = 24.71\pm 0.04$ (stat) $\pm 0.01$ (sys)~mag, 
$\mu_0 ~(TRGB_I) = 24.72 \pm 0.02$ (stat) $\pm0.07$ (sys)~mag, 
$\mu_0 ~(TRGB_{NIR}) = 24.72 \pm 0.04$ (stat) $\pm0.10$ (sys)~mag, and 
$\mu_0 ~(JAGB)=24.67\pm0.03$ (stat) $\pm0.04$ (sys)~mag. We also measured a distance modulus to M33 based on the JAGB method using HST photometry from the PHATTER survey, measuring $\mu_0=24.71\pm0.06$ (stat) $\pm~0.05$ (sys)~mag. We showed that selecting JAGB stars using the color ($F814W-F110W$) is suitable for future space-based JAGB studies in two ways: (1) Theoretically, using stellar isochrones and (2) Empirically, as the values determined using ground-based and space-based agreed well.

Currently the Leavitt law and TRGB provide the most robust local determinations of $H_0$, each with the highest number of SN Ia calibrators and both having been well-scrutinized and dissected for systematics. The two methods agree very well for nearby galaxies ($<10$~Mpc); \cite{2019ApJ...882...34F} found the scatter in a galaxy-to-galaxy comparison for the TRGB and Cepheid distances to be 2\%. However, the two methods begin to diverge in their estimations of greater distances, explaining the 2-$\sigma$ disagreement in their respective computations of the Hubble constant. The JAGB method may help to reveal hidden systematics or evidence for new physics in either or both methods. 

We also reiterate here that the methods for measuring distances based on the TRGB, Cepheids, and JAGB stars are completely independent. The RGB stars are an older, metal-poor population, found in the halos of galaxies. Cepheids are young, metal-rich stars found in the star-forming disks. JAGB stars are intermediate-age stars that can be found in the outer disks. The mechanisms by which they are effective distance indicators are also entirely separate; for the TRGB: the helium flash, for Cepheids: mechanical pulsation cycles; and for the JAGB stars, thermal instabilities that lead to dredge-up episodes. Thus, many of the systematics that afflict each distance indicator are independent and can be unearthed through inter-comparison.

We emphasize that the data used in this study were individually optimized for each independent distance indicator measurement. The TRGB measurements (both I-band and NIR) were analyzed in the outer regions of M33 to avoid the crowded, higher-reddening parts of the disk. The Cepheid measurement was performed in the least-crowded star-forming spiral arms in M33. The JAGB measurements, both space-based and ground-based, were performed in the outer disk of M33, where crowding and reddening are both reduced, but where intermediate-age AGB stars still reside.  We find the measured distance moduli in this study are in excellent (2\%) agreement, likely a result of our deliberate imaging selections. 

The upcoming launch of JWST will provide high-quality data for a variety of distance indicators, but will be incredibly advantageous for the JAGB method in particular. NIRcam boasts increased angular resolution in the near-infrared ($0.07\arcsec$/pixel) over HST's WFC3/IR camera ($0.13\arcsec$/pixel) which will help with blending/crowding effects. NIRcam also possesses the ideal filter combination (F115W$-$F150W) to distinguish the JAGB stars in color-magnitude space. With an expected reach of 100 Mpc using JWST (the current distance limit of the TRGB and Leavitt law with HST is about $\sim40$~Mpc), the JAGB method will have the ability to observe significantly more SN Ia calibrators and even provide a direct measurement of $H_0$ in the more distant Hubble flow itself.

\acknowledgments
We thank  David Bersier, Phil Massey and Vicky Scowcroft for generously providing us with their M33 catalogs. We also thank the anonymous referee for their constructive suggestions.

AJL would like to recognize that the UKIRT and CFHT astronomical observations described in this paper were only possible because of the dispossession of Maunakea from K'anaka Maoli, and stands in solidarity with the Pu'uhonua o Pu'uhuluhulu Maunakea in their effort to preserve this sacred space for native Hawai'ians.

AJL thanks the LSSTC Data Science Fellowship Program, which is funded by LSSTC, NSF Cybertraining Grant \#1829740, the Brinson Foundation, and the Moore Foundation; her participation in the program has benefited this work. MRC acknowledges support from the European Research Council (ERC) under the European Union’s Horizon 2020 research and innovation programme (grant agreement no.682115).

Some of the data presented were obtained with the WIYN 3.5-m telescope, Kitt Peak National Observatory, National Optical Astronomy Observatories, which is operated by the Association of Universities for Research in Astronomy, Inc. (AURA) under cooperative agreement with the National Science Foundation. The WIYN Observatory is a joint facility of the University of Wisconsin-Madison, Indiana University, Yale University, and the National Optical Astronomy Observatories. 
Some of the data presented were obtained at the Canada-France-Hawaii Telescope (CFHT) which is operated by the National Research Council (NRC) of Canada, the Institut National des Science de l'Univers of the Centre National de la Recherche Scientifique (CNRS) of France, and the University of Hawaii. 
Some of the data presented were obtained at UKIRT, which is owned by the University of Hawaii (UH) and operated by the UH Institute for Astronomy; operations are enabled through the cooperation of the East Asian Observatory.
This research has made use of the NASA/IPAC Extragalactic Database (NED) and the NASA/IPAC infrared Science Archive (IRSA), both of which are operated by the Jet Propulsion Laboratory, California Institute of Technology, under contract with the National Aeronautics and Space Administration. 
This publication makes use of data products from the Two Micron All Sky Survey, which is a joint project of the University of Massachusetts and the Infrared Processing and Analysis Center/California Institute of Technology, funded by the National Aeronautics and Space Administration and the National Science Foundation.
This research is based on observations made with the NASA/ESA Hubble Space Telescope obtained from the Space Telescope Science Institute, which is operated by the Association of Universities for Research in Astronomy, Inc., under NASA contract NAS 5–26555. These observations are associated with program \#GO-14610. WLF is partially supported under program \#GO-13691.
The Digitized Sky Surveys were produced at the Space Telescope Science Institute under U.S. Government grant NAG W-2166. The images of these surveys are based on photographic data obtained using the Oschin Schmidt Telescope on Palomar Mountain and the UK Schmidt Telescope. The plates were processed into the present compressed digital form with the permission of these institutions.
This research made use of the cross-match service provided by CDS, Strasbourg. This research has made use of NASA's Astrophysics Data System Bibliographic Services.

Finally, we thank the {\it Observatories of the Carnegie Institution for
Science} and the {\it University of Chicago} for their support of our long-term research into the calibration and determination of the expansion rate of the Universe. 

\facility{UKIRT (WFCAM, UIST, UFTI), CFHT (WIRCam), WIYN (S2KB, MiMo), HST (ACS/WFC, WFC3/IR)}

\software{ TOPCAT \citep{2005ASPC..347...29T}, Astropy \citep{2013A&A...558A..33A, 2018AJ....156..123A}, NumPy \citep{2011CSE....13b..22V}, Matplotlib \citep{2007CSE.....9...90H}, scipy \citep{2020NatMe..17..261V}, Pandas \citep{pandas} }

\end{document}